\documentclass{eas}
\usepackage{graphicx,natbib}
\begin{document}

%%-----------------------------
%%      the top matter
%%-----------------------------
\title{Stellar variability in radial velocity} 
\author{N. Meunier}\address{Univ. Grenoble Alpes, CNRS, IPAG, F-38000 Grenoble, France}
%\author{...}\address{...}
%\author{...}\address{...}
%
%
\begin{abstract}
Stellar activity due to different processes (magnetic activity, photospheric flows) affects the measurement of radial velocities (RV). Radial velocities have been widely used to detect exoplanets, although the stellar signal significantly impacts the detection and characterisation performance, especially for low mass planets. On the other hand, RV time series are also very rich in information on stellar processes. In this lecture, I  review the context of RV observations, describe how radial velocities are measured, and the properties of typical observations. I  present the challenges represented by stellar activity  for exoplanet studies, and describe the processes at play. Finally, I  review the approaches which have been developed, including observations and simulations, as well as solar and stellar comparisons. 
\end{abstract}
\maketitle
%%-----------------------------
%%      your text
%%-----------------------------
\section{Introduction}

Stellar activity impacts the measurement of radial velocities (hereafter RV), which in turn affects the detectability of exoplanets using this indirect technique: this was recognized very early-on \cite[][]{saar97} after the detection of 51Peg~b \cite[][]{mayor95}. This lecture therefore focuses on integrated RV observed for stars other than the Sun, and on the effect of both magnetic activity (due to spots and plages) and variability due to photospheric dynamics at various spatial scales (from granulation to large scale flows), as well as the interaction between these two categories of processes. Asteroseismology as well as Doppler imaging are outside the scope of this lecture. We focus on F-M main sequence stars, with a bias toward old solar-type stars. 

%test1
%\cite{saar97} pour Saar (1997)
%\citep{saar97} pour (saar xxx 1997)
%\cite[][]{saar97} pour entre () (saar etXXX, 1997)

\begin{figure} 
\includegraphics[width=12.5cm]{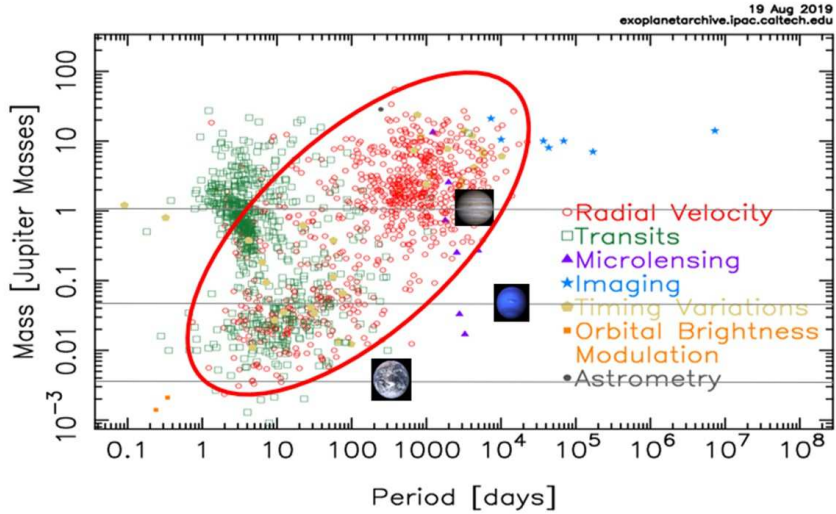} 
\caption{Planet mass vs. planet period (projected mass when observed with radial velocity) for different detection techniques. Jupiter, Neptune and the Earth are indicated for comparison. }
\label{exopla} 
\end{figure}

RV is an indirect technique in exoplanet studies, i.e. we observe the light coming from the star and not from the planet (Fig.~\ref{exopla}): if the star impacts the RV measurement (by modifying the line shape or its position), then exoplanet detectability and characterization are also affected. This is true for several indirect  techniques (RV, photometry, astrometry), which are complementary in terms of the orbital range they cover, but also in terms of the information they provide (for example, mass is obtained with radial velocities and radius with transit photometry, giving clues on the planet density). 
However, although RV are very sensitive to stellar activity, they are not the traditional way to study it: RV observations and surveys are usually biased toward the search for exoplanets (studies in which stellar activity is usually considered as ``noise"), while several other methods are commonly used to study stellar activity such as chromospheric emission, photometry, or spectro-polarimetry. 

\begin{figure} 
\includegraphics[width=12.5cm]{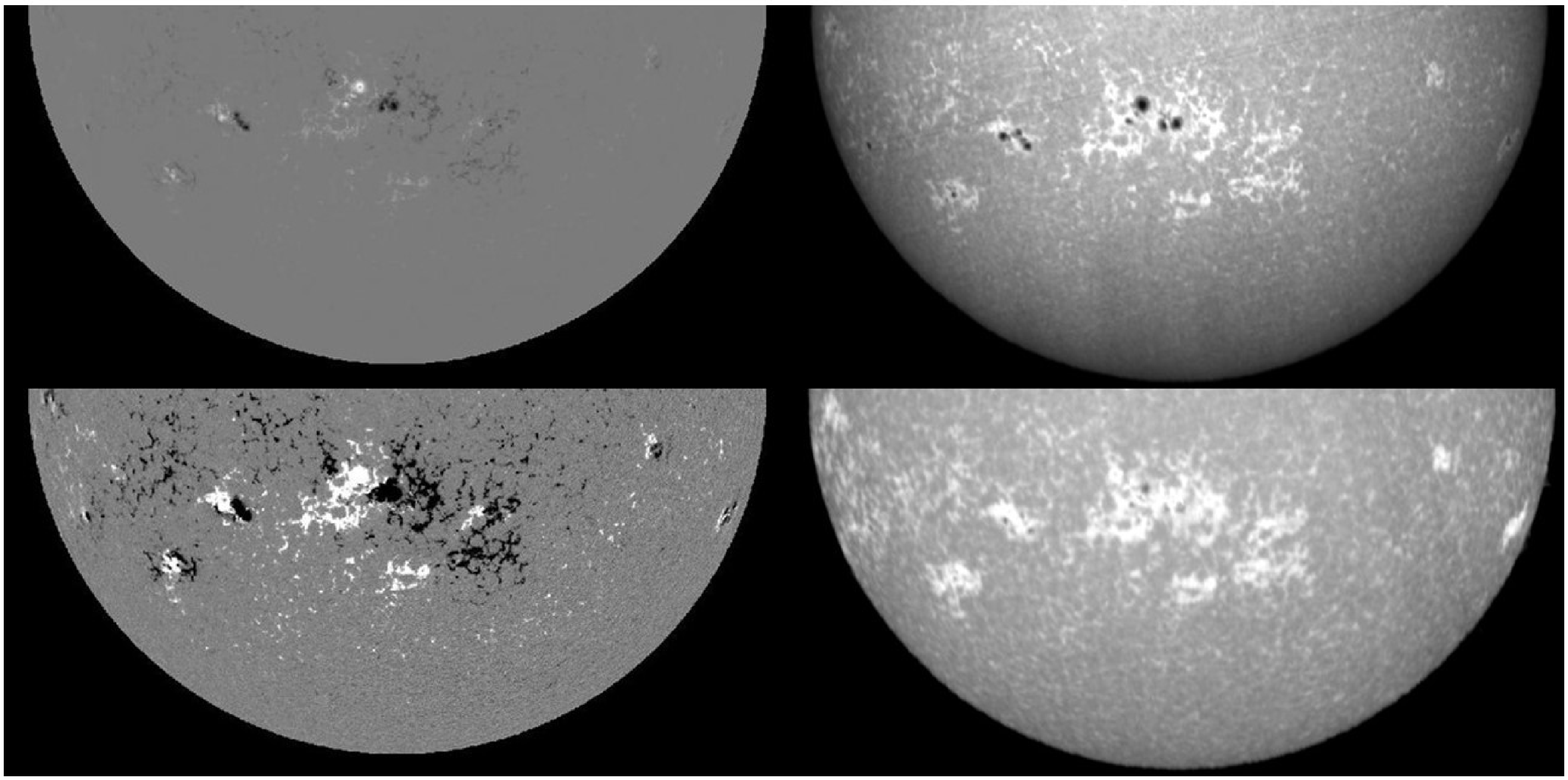} 
\caption{Solar magnetogram obtained by MDI/SOHO (left, with two different magnetic field saturation on the figure), and images of the solar photosphere (upper right panel) and chromosphere (lower right panel) from Meudon Observatory (BASS2000). } 
\label{degener}
\end{figure}

Furthermore, most observables used to study stellar activity are strongly degenerate, as illustrated in Fig~\ref{degener}: for example, there is a strong degeneracy between spots and plages or between structure size and contrast when analyzing photometric light curves, while polarimetry is very sensitive to flux cancellation. RV measurements are no exception in that respect: they are sensitive to many processes, some of them subject to degeneracies. 
However, all these techniques are sensitive to different processes and are therefore complementary from the point of view of stellar activity. Combining different approaches and activity indicators is therefore very useful to obtain a complete view of stellar variability for such stars: RV in particular are sensitive to several processes not affecting other observables, which is a problem from the point of view of exoplanet detection, but also very interesting to study stellar processes related to magnetic activity or photospheric flows. Studying stellar activity using RV methods is also necessary to push the limits of exoplanet detections. Finally, a great wealth of knowledge has been obtained for the Sun: solar physicists are used to exploit the very high spatial resolution at our disposal, often without considering the integrated signal (except for  irradiance and some helioseismology observations mostly), but leading to results of great interest to understand stellar RV. 
The content of this lecture therefore lies at the interface between stellar physics, solar physics and exoplanet studies.

The outline of the lecture is the following. I first introduce the context, how radial velocities are measured and affected, and what are the current challenges. Then I review the stellar processes impacting radial velocity measurements in detail and provide information about their typical amplitude and timescales. Finally, I describe the approaches which have been developed by different groups to deal with the challenge consisting in disentangling the planetary contribution from the stellar one: stellar and solar observations, correction techniques and simulations. I conclude with some perspectives for the next few years.

\section{From the instruments to radial velocities}

The RV technique needs stabilized spectrographs associated to high spectral resolution to be able to measure very precise RV on a long-term basis ($>$ 10 years). Very stable instruments have been developed and implemented to be able to detect exoplanets, which are inducing  a reflex motion of the star: this technique, based on the measurement of the position of spectral lines using various techniques, led to the first detection of an exoplanet around a main-sequence star in 1995 \cite[][]{mayor95}. I present here a short overview of instruments and observations. 

\subsection{Instruments}

Such spectrographs have been developed for the purpose of detecting exoplanets using RV techniques (this list is not exhaustive as many facilities exist) in two wavelength domains: first in the visible (e.g. Sophie at the 1.93m, Observatoire de Haute Provence, HARPS at the 3.6m in La Silla, HARPS-North at the TNG in La Palma, Espresso at the VLT in Paranal using one of the 8m UT or the 4 telescopes) and more recently covering the infrared domain as well (e.g. Carmenes at the Calar Alto Observatory, Spirou at the CFHT, Hawaii, or the soon to come NIRPS in La Silla). Their main characteristics are: a high resolving power (in the 70k-190k), a very high long-term stability (with goals between 10cm/s and 1 m/s depending on the instrument and wavelength domain), and a good signal-to-noise ratio. They are characterized by their transmission, wavelength domain, and sampling. The S/N ratio on the spectrum depends on the spectral type and magnitude of the star, and on the order in the spectrum, but is typically in the 10-500 range. Reaching values down to 10 cm/s constitutes a huge instrumental challenge, as a displacement of the spectrum on a typical CCD detector represents a few atoms only! 

\subsection{From spectra to radial velocities}

These instruments produce Echelle spectra, each of them typically recording the stellar spectrum as well as a reference  spectrum used for the precise wavelength calibration (Th lines, Fabry-Perot etalon, ...) for each order (for example 72 for HARPS). RVs are usually computed for each order separately and then combined. Different methods have been implemented to measure RV from these spectra. The most common one (for example used in the standard SOPHIE or HARPS pipeline) is based on a cross-correlation between the observed  spectrum and a binary mask indicating the position of the expected lines for the spectral type of the observed star. The position of the maximum of this cross-correlation function (CCF), found by adjusting  a symmetric Gaussian function, provides the stellar RV for this observation. Several groups have also developed complementary techniques, which are more suitable for certain types of stars, and in which the correlation function is computed with a reference spectrum instead of a binary mask, usually a median spectra derived from the whole time series of spectra for that star. This has been done in the Fourier space \cite[][]{chelli00,galland05} or in the temporal space \cite[][]{angladaescude12,astudillo15,astudillo17b}, to adapt the algorithm to very massive stars with very few broad spectral lines and to M stars with many blended lines respectively. There are also prospects to compute the RV more globally using Gaussian Processes \cite[][]{rajpaul20}. A few typical amplitudes give an idea of the difficulty of this task. Solar line widths are typically 2 km/s for example, and  faster rotators have line widths higher than  100 km/s, while a HARPS pixel is about 0.01~\AA$\:$ (corresponding to ~600 m/s at 5000~\AA). For comparison, detected planets have amplitudes of the order of the m/s up to a few 100 m/s for the most massive ones, while the Earth is below 10 cm/s only. 

\subsection{Observational strategies}

Because RV surveys have been biased toward the search for exoplanets, the temporal sampling is not necessarily optimal for stellar activity. The sampling is usually very irregular, with  a bad phase coverage of the stellar processes (for example of the rotational period). The sampling also affects the typical orbital periods which can be detected. This sampling results from various observational constraints such as telescope time allocation, the observability of the star, and the fact that  few nights per year are available to observe a given star (visibility of the star and reasonnable airmass to observe it when visible). Large surveys have been implemented, with data publicly available in some cases, for example on the ESO archive for HARPS: old FGK main sequence stars \cite[][]{mayor11,udry19}, old main sequence M stars \cite[][]{bonfils13,mignon20}, A-F stars \cite[][]{borgniet17,borgniet19}, young M-F stars \cite[][]{lagrange13,grandjean20}. These surveys are however often biased toward the less active stars in the considered category, with the objective to minimise the effect of stellar activity and maximise exoplanet detection rates.

\begin{figure} 
\includegraphics[width=9cm]{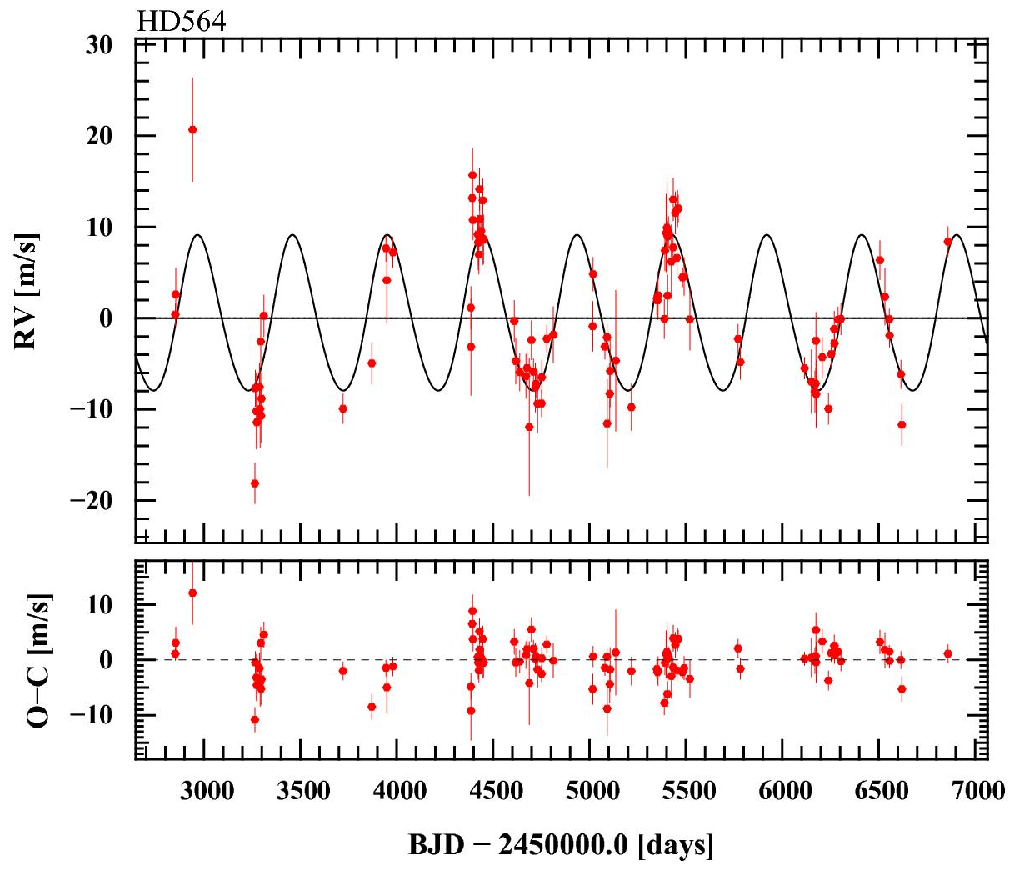} 
\includegraphics[width=9cm]{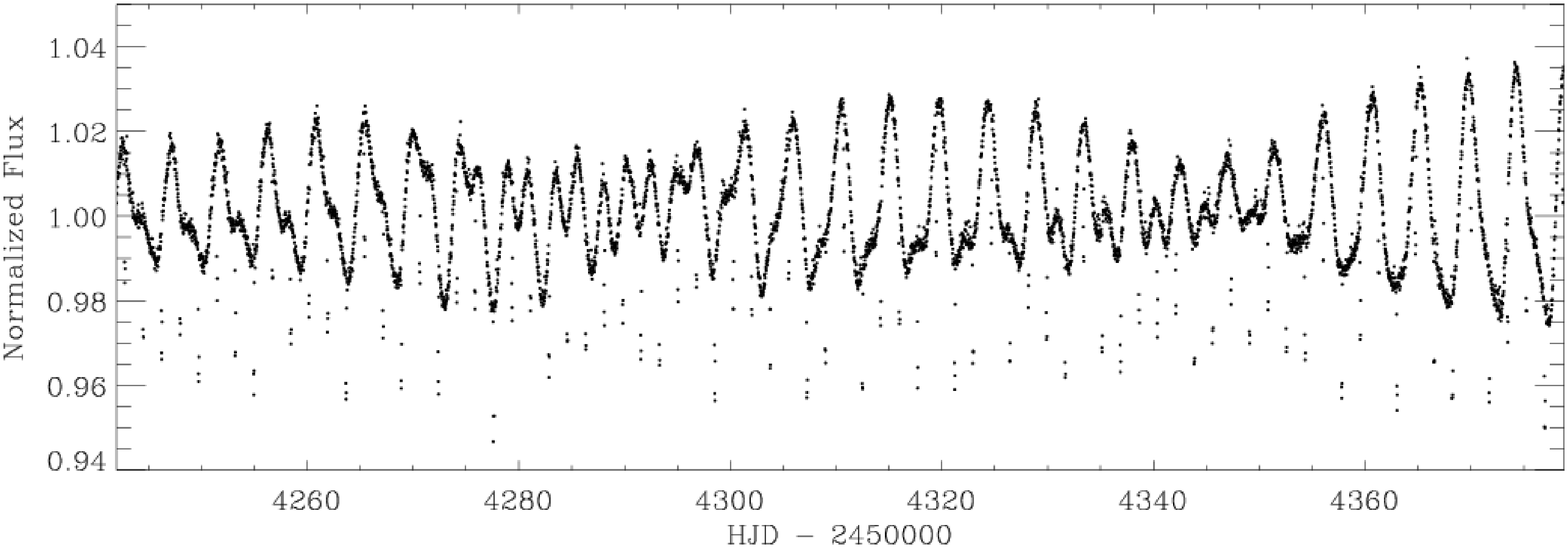} 
\caption{Upper panel: HARPS RV data for HD564, from \cite{moutou15} reproduced with permission \copyright ESO. Lower panel: example of a photometric time series with CoRot, showing the stellar variability superimposed to the transit of CoRot-2b, from \cite{alonso08} reproduced with permission \copyright ESO.  } 
\label{rv_phot} 
\end{figure}

RV technique applied to exoplanet detection is not the only one to be affected by stellar activity, since this is also the case for photometric transits. However, although the stellar signal is present at all time in both cases, the planet contributes to the signal in a very different manner: it is always present in the case of RV, while this is the case only during the planetary transits for photometry, meaning that in the latter case, the light curves are most of the time free of the planetary signal, as shown in Fig.~\ref{rv_phot}. It is therefore much easier to disentangle the planet signal from the stellar signal in this case (the transit depth can however be impacted by stellar activity, see lecture by G. Bruno, this volume). In addition, photometric light curves (from spatial missions such as CoRot, Kepler, or TESS) are always very well sampled, while RV samplings are extremely irregular and sparse, as illustrated in Fig.~\ref{rv_phot} for two typical time series.

Stellar activity also affects the astrometric signal, but for massive planets to be detected with Gaia, the amplitude of the effect is completely negligible for solar-type stars. Future high-precision astrometric mission aiming to detect very low mass terrestrial planets will be sensitive to stellar activity however, but to a much lesser extent than radial velocities for similar stars and planets \cite[][]{makarov10,lagrange11,meunier20}.

\subsection{Complementary information: chromospheric emission}

When searching for exoplanets, observables which are sensitive to stellar activity but not to the presence of a planet are routinely produced in order to help disentangling a possible planetary contribution  from the stellar signal. They are computed from the same spectra used to obtain  RV (and are therefore simultaneous): FWHM (full width at half maximum) and quantification of the line shape variability (for example the BIS, defined as the difference between the positions of the bisector at two different levels). Another very widely used indicator is a measure of  the chromospheric emission, strongly related to magnetic activity, such as the $\log R'_{HK}$ indicator, representing the emission in the Ca II H and K lines (at 3933 and 3968~\AA): a first indicator is defined by the emission integrated over the center of the line normalized by the continuum (S-index). The photospheric contribution is then subtracted, using a calibration depending on B-V in the 0.6-1.2 range \cite[][]{noyes84} and more recently for M stars \cite[][]{astudillo17}. This calibrated flux is then corrected by the bolometric flux \cite[][]{noyes84} to allow to compare stars of very different spectral types, which leads to the routinely produced $\log R'_{HK}$ indicator. The estimate of this indicator is likely to present  uncertainties due to these calibrations, of the order of 0.05 for the average level of a given star \cite[][]{radick18}, but its variability is more precise and very useful to identify the presence of magnetic activity. The chromospheric emission is strongly related to plages, i.e. bright magnetic area in the photosphere (and not to dark spots), as can be seen in solar images taken in the same wavelength band (BASS2000, http;//www.bass2000.obspm.fr/), as shown in Fig~\ref{degener} (lower right panel), or for example the high spatial observation around a small solar structure \cite[][]{grant15}. The chromospheric emission can also be derived from other chromospheric lines, for example H$\alpha$, but the correlation with the $\log R'_{HK}$ (as a function of time) is not always good \cite[][]{cincunegui07}, which could be due to stellar processes \cite[][]{meunier09}.

\subsection{Effect of stellar variability on exoplanet studies in radial velocity: impact and challenges}

\begin{figure} 
\includegraphics[width=12.5cm]{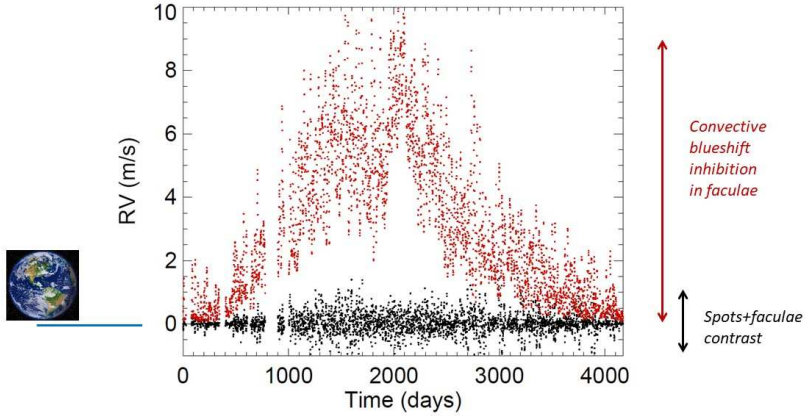} 
\caption{Reconstruction of the solar integrated RV due to magnetic regions, due to the spot and plage contrasts (black) and including the inhibition of the convective blueshift in plages (red) made by \cite{meunier10a}, compared with a Earth signal ($\sim$ 9 cm/s, smaller than the thickness of the blue line).} 
\label{challenge} 
\end{figure}

Stellar variability, due to various processes (mostly magnetism and photospheric flows as well as processes related to both categories), varies at different timescales: they will be detailed in the next section. Their impact on exoplanet observations can take different forms:

\begin{itemize}
\item{{\it It can look like an exoplanet}, and in the past led to exoplanet detection publications which were later invalidated. For example, the re-analysis \cite[][]{robertson14} of the four planets detected around Gl581 \cite[][]{udry07,mayor09} showed that planet d did not exist, and that the observed signal at its period was due to stellar activity. Another example is the publication of two Super Earths in resonance orbiting HD41248 using 62 points of public data \cite[][]{jenkins13}, while the complete analysis by the PI of the observations \cite[][]{santos14} on many more points showed that one of the detections was due to stellar activity, while the other one did not exist. Most of recent publications take the presence of stellar activity  carefully into account, using activity indicators as described in Sect. 2.4, although all mitigation techniques have limitations. This will be discussed in Sect.~4. The estimation of the false positive level is however a difficult issue \cite[][]{sulis17}. }
\item{{\it It can hide an exoplanet} because of the strong additional signal. As an example, Fig.~\ref{challenge} shows the reconstruction   of the solar RV signal during cycle 23 \cite[][]{meunier10a}, leading to the conclusion  that the inhibition of the convective blueshift in plages is dominating over the contribution due to the contrast of spots and plages: it has a long term amplitude of about 8 m/s, which is two orders of magnitude higher than the Earth signal, hence a huge challenge. }
\item{{\it It can affect planet characterisation}, although this has been less studied. The estimation of the planet mass with  RV techniques in transit follow-ups can therefore be noisier or biased. The presence of the stellar signal  leads to significant uncertainties on the masses determined by RVs and possibly to some biases, which in turn will impact the estimation of the planet densities. It could also affect the Rossiter-McLaughlin effects, or  planetary atmosphere characterization. }
\end{itemize}

\section{Stellar processes contributing to radial velocities}

\subsection{General overview}

\begin{figure} 
\includegraphics[width=12.5cm]{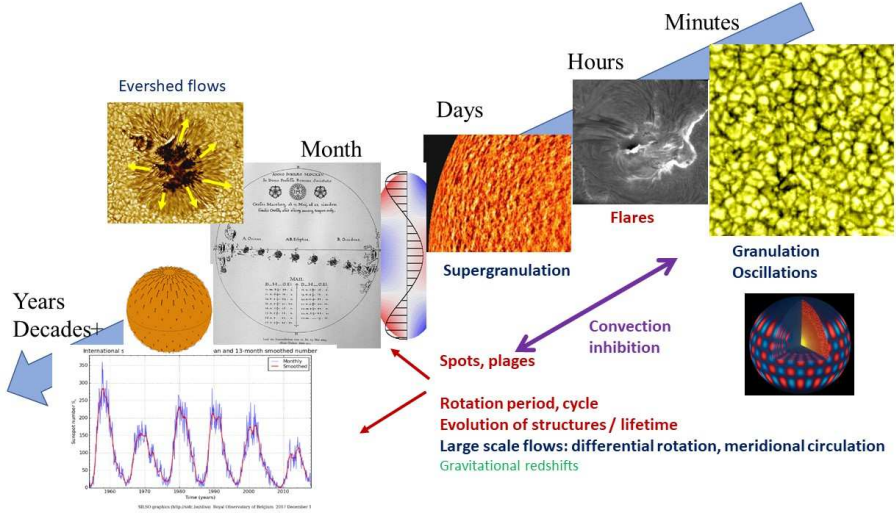} 
\caption{Overview of the different processes at play, due to magnetic regions (red), flows (blue), interaction between the two (purple), and gravitational redshift (green). Typical time scales correspond to the Sun. Granulation image: Pic du Midi Observatory; Flare: H$\alpha$ image, Big Bear Solar Observatory; Supergranulation: Dopplergram MDI/SOHO; oscillations: example of a mode from the GONG project, NSO; differential rotation: NASA/Marshall Solar Physics; spot drawings: Scheiner, 1625; Evershed flows: photosphere image from Vacuum Tower Telescope, NSO/NOAO; meridional circulation flows: Fig. 1 from \cite[][]{makarov10b}, reproduced by permission of the AAS; spot number vs. time from SIDC.  } 
\label{panorama} 
\end{figure}

Stellar activity varies on different time scales. Figure \ref{panorama} shows a summary of the different processes at play, where the indicated scales are indicative of the typical solar case: they may be quite different for other spectral types or young stars. These processes are due to magnetic structures, to flows (in particular in the photosphere), to interaction between magnetism and flows, and to gravitational redshifts. At the shortest timescales, oscillations and granulation play an important role, although granulation power extends to much lower frequencies, so that the notion of typical time scales is in fact ill-defined. They are followed by supergranulation, with a lifetime of the order of 1-2 days, again with a power spectrum extending to very low frequencies, including in the habitable zone range. Flares are sporadic features, with duration typically below one hour, often down to a few minutes. Spots and plages have a strong effect for periods close to the rotation period, with a complex power spectrum due to the presence of differential rotation and the finite lifetime of these structures,  an amplitude which may vary over time (i.e. a modulation of the amplitude on long timescales, for example during the cycle due to the dynamo process, again showing that the notion of timescales is not well defined), and the presence of harmonics. The inhibition of the convective blueshift in plages is also rotationally modulated but has in addition a stronger impact at long timescales (cycle). The gravitational redshift could be due for example to stellar radius variation with the cycle, but the expected amplitude is very small \cite[][]{cegla12}. In the following, I review in more details most of these processes. The Sun will often be used as a reference in this paper. I  will provide the corresponding  timescales and spatial scales, and describe how they are expected to depend on spectral type. 

\subsection{Spot and plage contrasts}

\begin{figure} 
\includegraphics[width=13cm]{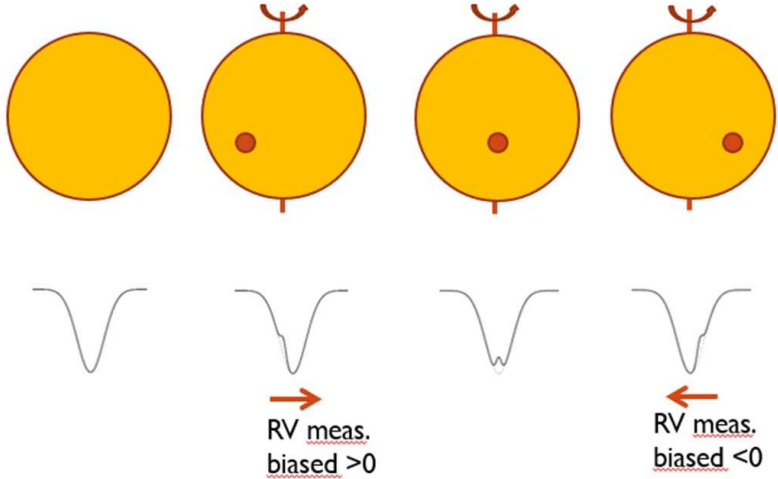} 
\caption{Illustration of the line distorsion due to a dark spot and its impact on the RV estimation. } 
\label{spot} 
\end{figure}

When rotating across the stellar disk, (dark) spots and (bright) plages change the flux coming from areas of the surface corresponding to different Doppler shifts (due to rotation), i.e. from blueshifted regions to redshifted regions, as illustrated in Fig.~\ref{spot}. As a consequence, the line is distorted at different positions in the spectral line with time (the same principle is used in Doppler imaging or Zeeman-Doppler imaging techniques, when the high rotation rate allows to use this information to reconstruct low resolution maps\footnote{See for example http://www.astro.uu.se/$\sim$oleg/
 and http://www.astronomy.ohio-state.edu/$\sim$johnson.7240/$\#$tomographygallery
.}). When measuring the RV from the line profile (or the CCF), the RV is then biased: for example, in the case of a dark spot, RV varies from a redshift to a blueshift, crossing a velocity of 0 when the spot is on the central meridian. It is reversed for plages (with a slightly different shape due to the position-dependent contrast of plages). When several structures at different longitudes are present and visible at the same time, the contributions from all these spots are therefore partially cancelling each other, producing a complex variability. The resulting RV is therefore a residual between the different contributions, which may lead to degeneracies when attempting to fit this signal.

\begin{figure} 
\includegraphics[width=9cm]{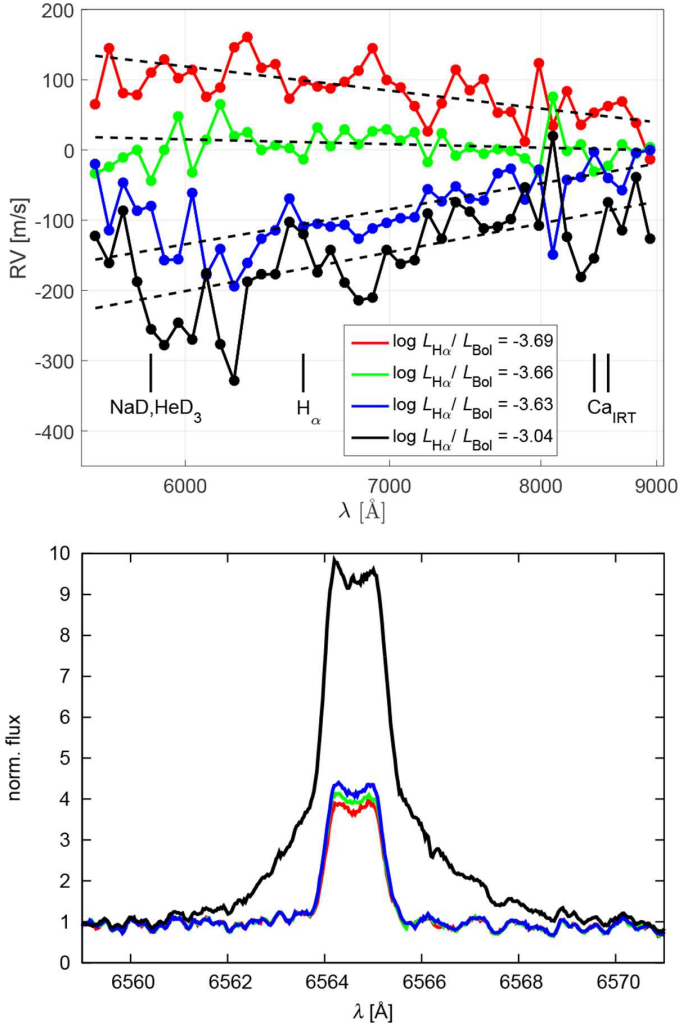} 
\caption{Upper panel: RVs vs. wavelength range (order) for different activity levels (the black one corresponds to a flare), the dashed line indicating the best-fit straight lines to each curve. Lower panel: corresponding H$\alpha$ emission.  From \cite{talor18} reproduced with permission \copyright ESO.  } 
\label{chroma} 
\end{figure}

In the solar case, the root-mean-square (hereafter rms) of the RV is of the order of 0.3-0.4 m/s   for both spots \cite[][]{lagrange10b}, plages \cite[][]{meunier10a}, and the sum of the two (due to partial cancellation). Peaks reach 1-2 m/s when the Sun is very active. The rms of the signal is higher during cycle maximum compared to cycle minimum. The signal is mostly present around the rotation period P$_{\rm rot}$ and at the P$_{\rm rot}$/2 harmonics. Due to the significant  differential rotation and the finite lifetime of structures, there are many peaks around the rotation period, covering a large range in period, and their amplitude strongly varies with time. The bisector also strongly varies with RV. RV are strongly affected by inclination \cite[][]{desort07}, since it is related to the projected rotational velocity, and by wavelength \cite[][]{talor18}, since it is a contribution due to the contrasts. The impact of wavelength is illustrated in Fig.~\ref{chroma}. The presence of strong magnetic fields also affects the measurements due to the Zeeman effect \cite[][]{reiners13}, which should be more important in IR compared to the optical.

For stars differing from the Sun in spectral types and ages, we expect spot and plage properties to differ from the solar ones. We first consider the properties of individual spots. A number of trends have been identified. First, the ratio between the plage and the spot filling factor (around 10 for the Sun) is expected to depend on age, with spots dominating for young stars \cite[][]{lockwood07,radick18}. Spot and plage modelling of photometric light curves of young stars also showed a plage-to-spot ratio of the order of unity \cite[e.g.][]{lanza09}. The spot temperature contrast, which is also an important parameter to determine the impact on the final RV, has been found to increase with increasing stellar effective temperature, from both observations \cite[][]{berd05} and recent models \cite[][]{panja20}, although they exhibit a large dispersion. The plage contrast is more complex because it depends on $\mu$ (cosine of the angle between the normal to the surface and the line-of-sight): the contrast is very low at disk center, and is increasing toward the limb, so that the plage contribution is much larger close to the limb. Models show that the contrast increases toward more massive stars, and also depends on magnetic field \cite[][]{norris16,norris18}. On the other hand, the size and lifetime of stellar spots and plages however is not known. Lifetimes are expected to increase toward less massive stars, because of the lower convection level \cite[leading to a weaker decay see][and references therein]{bradshaw14,giles17}, and a similar trend has been observed from Kepler data \cite[][]{giles17}. Furthermore, very long lived spots have been observed on M dwarfs, up to 1-2 yr, for example on GJ674 \cite[][]{bonfils07}. This is also the case for younger stars. The size dependence is not constrained however, because of the strong degeneracy between size and contrast, and between spots and plages. There may be some possibilities to derive such  trends from in-transit spot modeling of light curves in the future, because the degeneracies are less present \cite[e.g.][]{czesla09,silva-valio11}.

%XXXXX Chromatic dependence Fig.~\ref{chroma} example from \cite{talor18}
%legende d'origine = CARMENES VIS measurements of J22468+443 (EV Lac). Top: Order-by-order RVs from four representative observations. Red, green, and blue show three observations with close-to-median Hα emission. Black plots an observation with extremely high Hα emission, taken during a strong flare event (Fuhrmeister et al. 2018). The corresponding log LHα∕Lbol values are given in the inset. Dashed lines show the best-fit straight lines to the RV–order scatter plots of each observation. Individual-order RV uncertainties are not shown in the plot for clarity, but they are on the order of 10–30 m s−1, depending on the RV-information content and the S/N in each order. The markers at the bottom specify the locations of the indicated chromospheric emission lines. Bottom: Parts of the VIS spectra from these four observations, centered on the Hα line. The colors are the same as in the top panel.

\begin{figure} 
\includegraphics[width=11.5cm]{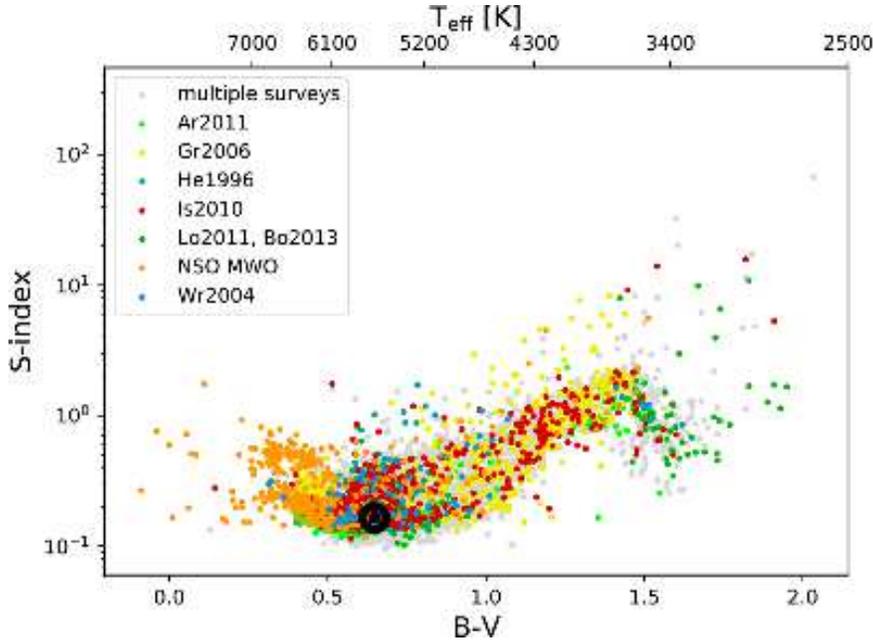} 
\caption{S-index vs. B-V of 4554 main-sequence stars, calibrated to the Mount Wilson scale. The black symbol represents the Sun during cycle minimum. From \cite{borosaikia18} reproduced with permission \copyright ESO.  } 
\label{logrphk} 
\end{figure}
%legende d'origine MWO (calibrated to the Mount Wilson scale) vs. B − V of 4454 main-sequence stars. The surveys they belong to are given in the legend (2509 are only listed in one survey, and 1945 appear in several surveys). The Sun at minimum activity is shown by the black ⊙ symbol.

\begin{figure} 
\includegraphics[width=9cm]{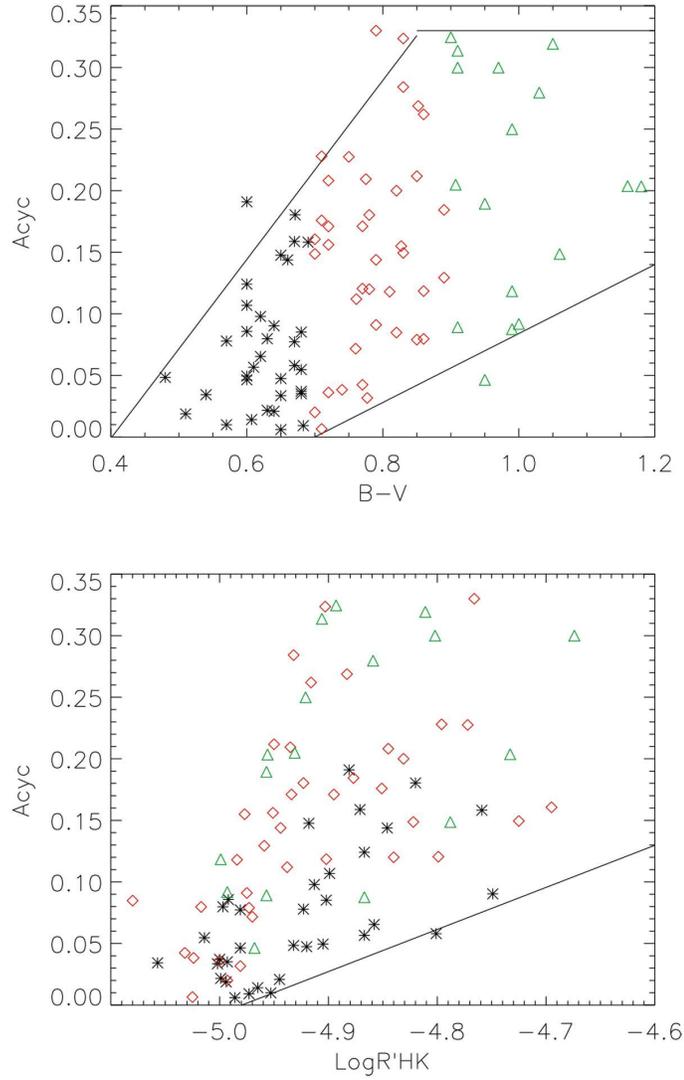} 
\caption{Semi-amplitude of stellar cycles Acyc (in 10$^5 R'_{HK}$) vs. B–V (upper panel) and vs. $\log R'_{HK}$ (lower panel), derived from Lovis et al. (2011) after revision of the largest amplitudes, for different types of stars: B–V $<$ 0.7 (black stars), 0.7 $<$ B–V $<$ 0.9 (red squares), and B–V $>$ 0.9 (green triangles). From \cite{meunier19} reproduced with permission \copyright ESO.  } 
\label{Acyc} 
\end{figure}

Another important consideration is the long term variability of stars, i.e. the modification in terms of activity level over time \cite[for example the presence of cycles or more stochastic variability, see Mount Wilson survey e.g. ][]{baliunas95} as well as the average level, related to the number of spots and plages. A useful observable to characterise these properties is the chromospheric emission (see Sect.~2.4). Most publications provide the average activity levels versus spectral type for a large sample of stars \cite[for example][]{gray03,gray06,mittag13}. Figure~\ref{logrphk} shows  recent results  \cite[][]{borosaikia18} for more than 4000 stars, which confirms the results from those earlier publications: for a given spectral type, a very large range of activity levels is covered (e.g. for G stars all activity levels are observed), and K stars show a lack of very quiet stars. The cycle amplitude also covers a large range for most spectral types, as shown in \cite{lovis11b} for example \cite[see also the adaptation of this relationship in][]{meunier19}, as illustrated in Fig.~\ref{Acyc}. 

%legende origine Upper panel: half full amplitude of stellar cycles vs. B–V, derived from Lovis et al. (2011) after revision of the largest amplitudes (see text) for different types of stars: B–V < 0.7 (black stars), 0.7 < B–V < 0.9 (red squares), and B–V > 0.9 (green triangles). The black lines correspond to the lower and upper boundaries that were taken into account in building the grid. Upper panel: same vs. average Log . The solid line represents the lower limit that was taken into account in building the grid.

\subsection{Oscillations}

Oscillations such as p-modes affect RV, with many peaks having periods of a few minutes and amplitudes of typically 1 m/s for the Sun. The large amount of peaks defines a well-defined envelope \cite[e.g.][]{kjeldsen95}. Because of their spectrum, they are easily averaged out \cite[][]{dumusque11b,chaplin19}. Their amplitude and frequency increase slightly with stellar effective temperature. In addition to those modes, some specific additional modes could be present: \cite{lanza19} have observed sectoral modes on the Sun, with the main mode having an amplitude of 0.44 m/s and a period of 16.19d (whose value depends on the rotation rate). For other stars, attention should therefore be given to the possibility for such modes to be present when attempting to detect low mass planets in this specific  period range. More massive, young stars can show much more intense pulsations, such as $\delta$ Scuti and $\gamma$ Dor, with longer timescales (minutes-hours or more). They can be of very high amplitude (km/s) and therefore critical to detect planets: this was for example the case for the detection of  $\beta$ Pictoris c \cite[][]{lagrange19}, since $\beta$ Pictoris  exhibits many modes with periods around 30 minutes \cite[][]{koen03,mekarnia17}.

\subsection{Granulation}

\begin{figure} 
\includegraphics[width=12cm]{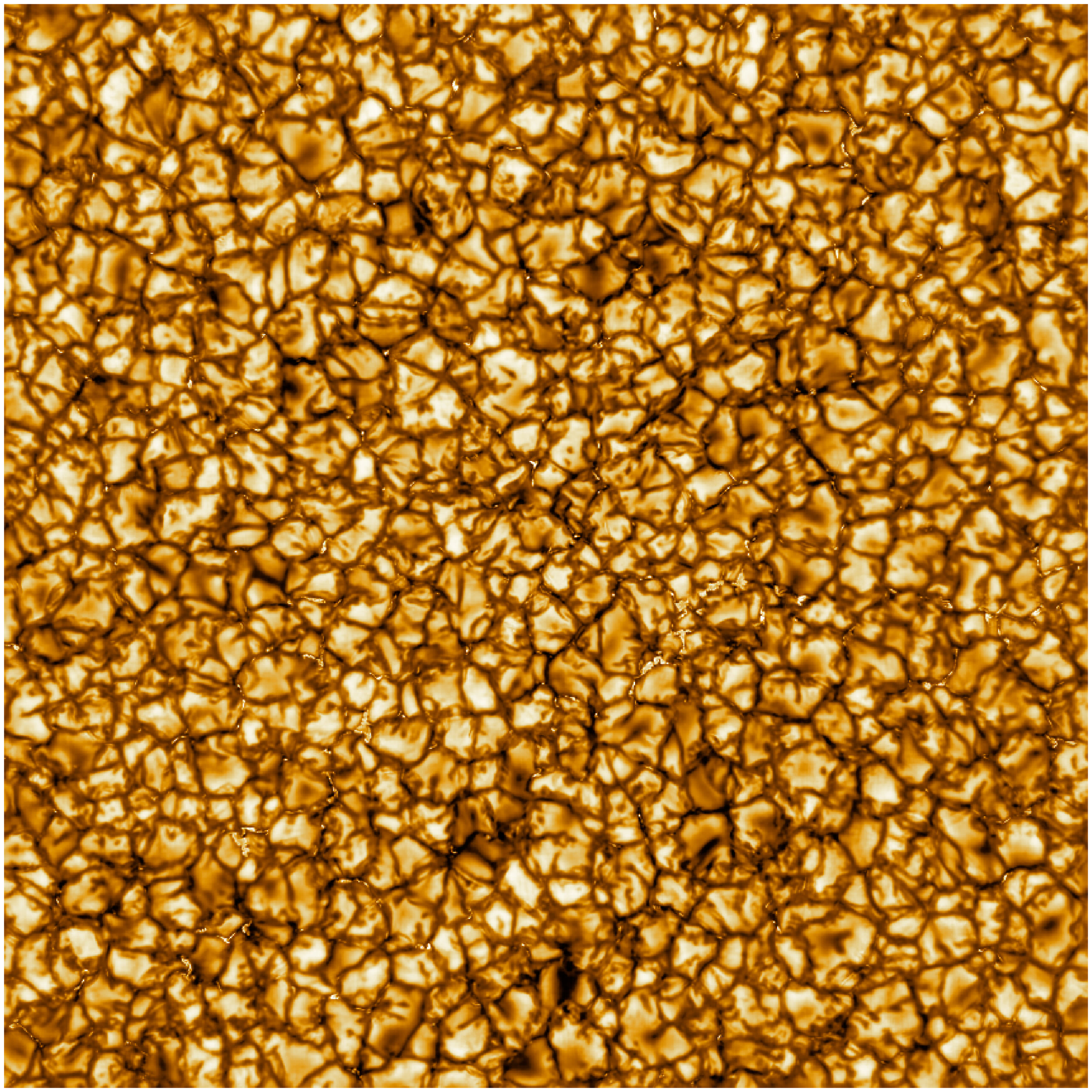} 
\caption{Image of solar granulation from the  Daniel K. Inouye Solar Telescope, taken at 789 nm, showing details of about 30 km. Credit NSO/AURA/NSF.  } 
\label{granulation} 
\end{figure}

%https://www.nso.edu/about/image-use-policy/
%legende origineCaption: The Daniel K. Inouye Solar Telescope has produced the highest resolution image of the Sun’s surface ever taken. In this picture taken at 789nm, we can see features as small as 30km (18 miles) in size for the first time ever. The image shows a pattern of turbulent, “boiling” gas that covers the entire sun. The cell-like structures – each about the size of Texas – are the signature of violent motions that transport heat from the inside of the sun to its surface. Hot solar material (plasma) rises in the bright centers of “cells,” cools off and then sinks below the surface in dark lanes in a process known as convection. In these dark lanes we can also see the tiny, bright markers of magnetic fields. Never before seen to this clarity, these bright specks are thought to channel energy up into the outer layers of the solar atmosphere called the corona. These bright spots may be at the core of why the solar corona is more than a million degrees!
%This image covers an area 36,500 × 36,500 km (22,600 × 22,600 miles or 51 × 51 arcseconds).

%%%%% rms of the smoothed granulation RV time series (red curve) versus scale. The rms RV of the corresponding residuals are in green. The horizontal dot-dashed line indicates a typical level of the noise induced by future instruments. 

Small scale convection (granulation) at the surface of the Sun has been identified and studied for a very long time, while solar images with unprecedented spatial resolution obtained with the recently built Daniel K. Inouye Solar telescope allows to access many details of the cell structures, as illustrated in Fig.~\ref{granulation}. Cells have a typical size of the order of 1 Mm and lifetime of 10 minutes, with a very large distribution of the values, leading to about 10$^6$ cells covering the visible disk. The flows are strong, of the order of 1 km/s (both vertically and horizontally). As a consequence, different realisations of those cells over the disk (i.e. at a given time) differ from one time to the other, so the average velocity is not equal to zero but exhibits a small residual which varies over time and is very stochastic. A recent review on the impact of granulation on RV  has been made by \cite{cegla19}. The typical rms (root-mean-square) of these residuals has been found around 0.4 m/s from specific lines \cite[][]{elsworth94,palle99}. A simulation made by \cite{meunier15} gave a larger rms (0.8 m/s), while the simulation of \cite{sulis20} based on numerical MHD simulations for the Sodium line as in the observations of \cite{palle99} provides a rms of 0.4 m/s in agreement with those observations. The simulations of \cite{cegla19} on the other hand leads to a low rms, about 0.1~m/s, which may be due to the choice of magnetic field used in those simulations.

\begin{figure} 
\includegraphics[width=9cm]{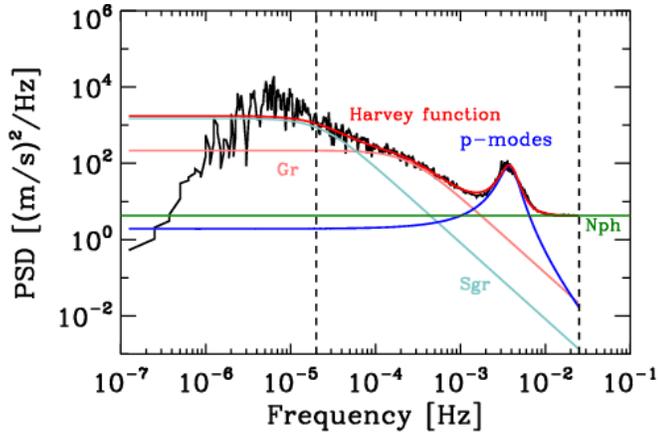} 
\caption{Solar power spectrum from GOLF/SOHO of a subset of $\sim$91 days, with a fit of the p-mode envelope (blue), supergranulation (blue-gray), granulation (pink), photon noise (green) and sum of the Harvey function for granulation and supergranulation (red). From \cite{lefebvre08} reproduced with permission \copyright ESO. } 
\label{power_gra} 
\end{figure}

%%%Figure 2: Results of two different type of fits, as explained in the text, applied to a GOLF spectrum of an arbitrary taken subseries of 91.25 days long. Top: left, PSD with a fit using 8 parameters (one lorentzian) to adjust the p-mode envelope; right, ratio between the PSD and the fit around the envelope of p-modes. Bottom: left, PSD with a fit using 11 parameters (two lorentzians) to adjust the p-mode envelope; right, ratio between thePSD and the fit around the envelope of p-modes. The dashed lines represent the limits inside which the fit is performed. The color used forthe different fits are: gray for the super granulation contribution, magenta for the granulation contribution, blue for the p-mode envelope, green for the noise and red for the harvey function (the sum of the granulation and supergranulation contributions).

\begin{figure} 
\includegraphics[width=12cm]{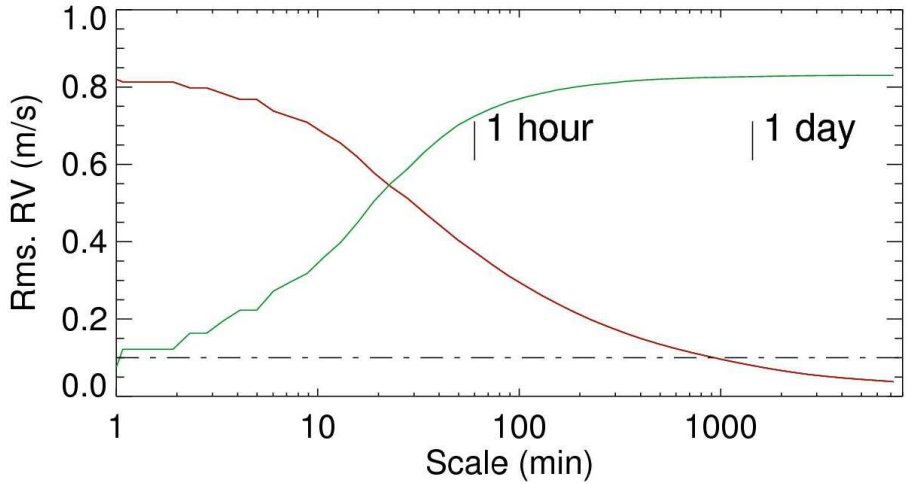} 
\caption{Rms of the smoothed granulation signal (in red) vs. the size of the binning box, while the rms of the residuals are in green. The horizontal dot-dashed line is the level expected from ESPRESSO/VLT. From \cite{meunier15} reproduced with permission \copyright ESO. } 
\label{rms_gra} 
\end{figure}

Granulation also strongly affects line shapes from solar observations  \cite[e.g.][]{dravins81} or from simulations \cite[e.g.][]{beeck13}, and the line shape variability can in turn represent  a useful indicator of the RV variability \cite[][]{cegla19}, although the amplitude seems to be small. 
The signal can be averaged to decrease its rms and its impact on exoplanet detectability \cite[][]{dumusque11b}. The signal is however very difficult to average out completely \cite[][]{meunier15}, because of the specific shape of the power spectrum.
The shape of the power spectrum of the granulation contribution has been modeled by \cite{harvey84} by a simple function, with a strong slope at high frequency, and a plateau at low frequency, meaning that they can significantly contribute to the power at all periods of interest for exoplanet searches. This is illustrated in Fig.~\ref{power_gra}. As a consequence, the rms as a function of the size of the temporal bin reaches an inflexion point around 1 hour, at which point the rms is divided by a factor of about two, then the decrease of the rms RV is extremely slow if we average over longer timescales (Fig.~\ref{power_gra}). Furthermore, the way the false positive are usually estimated may be biased \cite[][]{sulis17}, and \cite{sulis17b} proposed to  use a standardized periodogram to improve the analysis of this contribution. 

Granulation amplitude strongly depends on many stellar parameters, such as temperature and mass, but also log g and metallicity. The velocity field increases with effective  temperature from numerical simulations \cite[][]{allendeprieto13,magic13,magic14,trampedach13,tremblay13,beeck13,chiavassa18}. The same trend is observed from observations, using convective blueshift (see below) estimations \cite[][]{gray09,meunier17,meunier17b} and fits of the Harvey's function to observed power spectra for  a few stars \cite[][]{dumusque11b}.

\subsection{Supergranulation}

\begin{figure} 
\includegraphics[width=12.5cm]{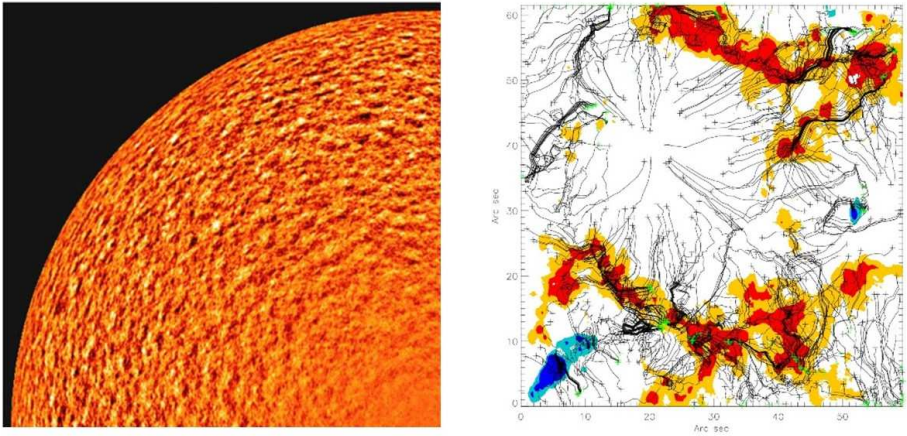} 
\caption{Left panel: Solar dopplergram from MDI/SOHO for a quarter of the solar disk, after averaging and subtraction of the rotation component, to exhibit supergranulation. Right panel: supergranulation flows and network concentration on the edge of a supergranule, from \cite{roudier16} reproduced with permission \copyright ESO. } 
\label{supergranulation} 
\end{figure}

In addition, flows at a larger scale than granulation are observed on the Sun, called supergranulation, also observed in the photosphere. Reviews on supergranulation can be found in \cite{rieutord10} and \cite{rincon18}. Supergranules are cells of the order of 30 Mm, with horizontal flows of the order of a few 100 m/s and very low vertical flows (below 30 m/s).These flows can be observed using Dopplergrams, as illustrated on the right panel of Fig.~\ref{supergranulation}. These flows advect the magnetic network toward the edges of the cells (left panel) and can therefore be observed using chromospheric emission maps or magnetograms exhibiting the network, or by tracking granules  to reconstruct the flows. The integrated RV is less well constrained than for granulation, but the principle is the same and due to the different realisations of the cell pattern over time. The flows are weaker, but there are also much fewer cells (a few thousands) and therefore we expect the residual to be important as well, while there is no photometric counterpart \cite[e.g.][]{meunier07c}. \cite{meunier15} obtained a range of 0.3-1.2 m/s for the Sun, while \cite{palle99} observed a rms of 0.78 m/s in a specific line. An example of time series comparing granulation and supergranulation following the Harvey laws are shown in Fig.~\ref{power_gra}. The resulting signal is much more difficult to average than granulation \cite[][]{meunier15,meunier19} because of the longer timescale involved: an average over 1 hour does not change the rms at all. \cite{meunier19,meunier20b} showed that supergranulation is more problematic to detect low mass planets (like the Earth) at long orbital periods (habitable zone), because of the amplitude and the low frequency power. 

It is difficult to quantify how supergranulation varies with spectral type, especially since the origin of supergranulation is not well understood. \cite{dumusque11b} performed fits of the Harvey function for a few stars observed by HARPS, but the sample is too small to derive a trend. \cite{meunier20} scaled it to  granulation amplitude because \cite{roudier16} showed that supergranulation appears  to be related to the behavior of granulation (through the analysis of trees of fragmenting granules). 

\subsection{Convective blueshift inhibition in plage}

Another effect of the presence of granulation is the convective blueshift \cite[e.g.][]{dravins81}, due to the different contributions to the integrated spectrum over the surface coming from the bright upflow areas of granules (large fraction of the surface, blueshifted) and from the dark downflow areas (small fraction of the surface, redshifted). The amplitude of this convective blueshift for the Sun is about 400 m/s \cite[see][for a recent estimation]{reiners16}. If the convective pattern is attenuated (with smaller weaker granules) by a different amount over time, for example  in magnetic plages, the resulting convective blueshift is modulated, with a net redshift at time when there are more plages compared to a quiet star. This leads to a modulation of the RV at the rotational period (as plages cross the disk) and on the long-term (during the activity cycle for example, since the signal depends on the amount of surface covered by plages). The simulations of \cite{meunier10a} showed that the resulting signal for the Sun over a complete cycle is the dominant contribution to the RV variation, with a long term amplitude of about 8 m/s. It also contributes significantly to the rotational modulation. The periodogram of the time series shows that it is responsible for a strong peak at the cycle period as expected, but also of large  power in the habitable zone, which is significantly higher than the power due to a low mass planet like the Earth. For the Sun, \cite{meunier10a} showed that the amplitude of the convective blueshift attenuation was much stronger in large plages (where the magnetic flux is higher) compared to smaller plages and small network features, with a ratio of about 6 between the maximum and the minimum \cite[and furthermore small features show a weaker cycle dependence, ][]{meunier05b}. Such a trend  was also observed later by \cite{milbourne19}. 

The convective blueshift has interesting properties which can be used to quantify this effect in other stars and also to develop mitigating technique. The most important one is that it depends on the line depth, with deep lines (formed higher in the atmosphere) exhibiting a much smaller convective blueshift than weak lines, formed higher in the atmosphere \cite[e.g.][]{dravins81,reiners16}. This property has been exploited by \cite{meunier17c} and \cite{dumusque18} to propose different mitigating techniques based on spectral line selection (See Sect.~5.1).  

Several applications of this property have been made in various papers to  stars other than the Sun \cite[][]{dravins87b,dravins89,allende99,landstreet07}. \cite{gray09} used this property to compare the convective properties of different stars: he computed the position (which is then converted into a velocity) of the bottom of selected lines in a 100~\AA$\:$ range as a function of the line depth and deduced that all stars showed a similar shape, i.e. a universal signature, which he called the third signature. Following this work, \cite{meunier17} and \cite{meunier17b} assumed a direct relation between the shape of the signal and the amplitude of the convective blueshift for a large sample of spectral lines and a large sample of stars: this allowed to confirm the trend of the amplitude of the convective blueshift with spectral type. It also allowed to observe a dependence of the convective blueshift on the average activity level (for a given spectral type), showing that it was indeed weaker in active stars: an attenuation factor (characterising the attenuation of the convective blueshift in plages) could be deduced from this analysis, with a possible trend with spectral type between 5400 and 6400 K. 

\subsection{Evershed flows}

On the Sun, outward horizontal flows have been observed around sunspots and characterised for a long time (upper left image in Fig.~\ref{panorama}). These flows are of the order of 1-2 km/s. If they are symmetric, a very small residual is expected after integration on the spot, but the resulting RV could be slightly higher, especially for large irregular spots. The effect has however not been quantified precisely \cite[see ][for discussions about photospheric flows associated to spots and active regions]{haywood20}.

\begin{figure} 
\includegraphics[width=8cm]{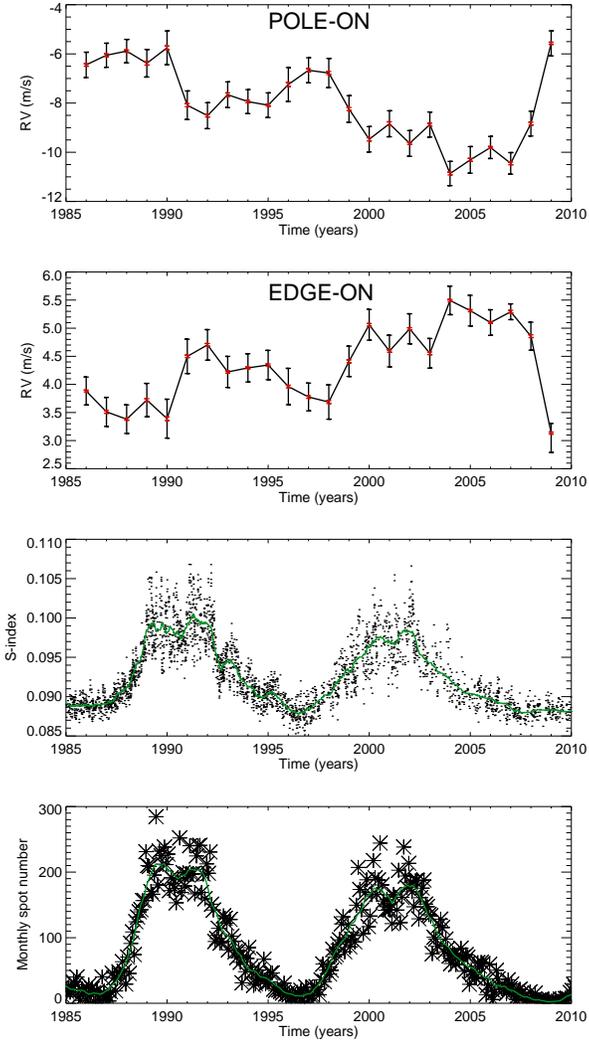} 
\caption{Solar integrated radial velocity due to meridional circulation reconstructed from the observed latitudinal profile of \cite{ulrich10}, over two solar cycles, for pole-on configuration (first panel), edge-on configuration (second panel), compared to the activity cycle from the chromospheric emission S-index (third panel) and spot number (fourth panel). From \cite{meunier20c} reproduced with permission \copyright ESO. } 
\label{fig_mc} 
\end{figure}
\subsection{Meridional circulation}

Meridional circulation constitutes a contribution to the integrated RV which has not attracted much attention compared to other processes. It is a large-scale flow which, if variable with time, should also lead to RV variability after integration on the disk. On the Sun, meridional circulation is a poleward flow, with a maximum amplitude in the 10-20 m/s range and varying with the solar cycle. It has been studied for several decades using various complementary techniques, such as Dopplergrams, magnetic feature tracking, or local helioseismology \cite[][]{duvall79,labonte82,howard86,ulrich88,komm93,snodgrass96,hathaway96,nesmeribes97,meunier99}. Converging flow patterns have also been observed in activity belts \cite[][]{meunier99,zhao04,lin18}. The first study estimating its impact on integrated RV has been made by \cite{makarov10b}, who performed a reconstruction of the solar variation, for the Sun seen edge-on: the resulting contribution was however superimposed to other large contributions and therefore its precise behavior was not clear. \cite{meunier20c} have therefore performed a new reconstruction, based on the solar measurements made by \cite{ulrich10} covering two solar cycles and for different inclinations between edge-on and pole-on. The resulting amplitude shows a significant variation with an amplitude over cycle 23 of the order of 2 m/s for the edge-on configuration and a few m/s for the Sun seen pole-on, with a reversal in sign between the two orientations for an inclination around 50$^\circ$. The results are shown in Fig.~\ref{fig_mc}.  The impact on exoplanet detectability is therefore important, especially for the extreme configurations, while intermediate inclinations may be the most suitable in that respect.

For other stars, the amplitude of meridional flows is decreasing for fast rotators \cite[][]{ballot07,brun17} and for low mass stars \cite[][]{matt11,brun17}. \cite{meunier20c} extrapolated the solar amplitudes of cycle 23 to a range of F6-K4 old main-sequence stars of various activity levels using the scaling laws from \cite{brun17} and assuming a proportionality to the cycle amplitude. They found an amplitude between 0.1~m/s for the most quiet stars and intermediate inclinations, up to 4 m/s for the most active stars.  The amplitude should be much smaller than those results for fast rotators if they exhibit a multicell pattern as expected from HD simulations \cite[][]{matt11,guerrero13,guerrero16} however.  

\subsection{Flares}

Flares produce a RV signal which is very stochastic and of very short duration. Because they are very localized on the surface and represent a small area, their impact is negligible for Sun-like stars. M dwarfs RV time series are often affected by flares however, because some of these stars are extremely active and exhibit very energetic flares: they appear as ``outliers" superimposed to the RV time series with amplitude of up to a few 100 m/s.

%%%%%%%%%%%%%%%%%%%%%%%%%%%%%%%%%%%%%%%%%%%%%%%%%%%%%%%%%

\section{Approaches to the problem}

As shown in the previous section, there are many sources of stellar variability impacting RV at various timescales. Many of them are in the range 0.3-1 m/s for the Sun, with a complex behavior for solar type stars, while the activity patterns may be more stable for young stars or M dwarfs. Such a complexity is due to the activity pattern itself, the evolution and finite lifetime of the structures, differential rotation, degeneracies between some contributions and complex temporal variability. None of those contributions are strictly periodic. In practice, the analysis of RV time series is also affected by the temporal sampling since it is most of the time sparse and irregular. To deal with this complexity, several complementary approaches have been developed by many groups. I will first briefly review the mitigating techniques which have been implemented. Then I will discuss  the approaches used to understand and quantify the contribution of the different processes to the RV measurements, and to test and improve mitigating techniques. 

\subsection{Mitigating techniques}

Given the very serious challenge represented by the presence of the stellar activity contribution to the RV time series when searching for exoplanets, many techniques have been implemented over the last 20 years. They all contribute to remove part of the stellar signal although none of them so far allows to reach very low levels of residuals, although a systematic analysis of the impact and performance is seldomly made. We list these approaches  in Table~\ref{tab_techn}. 

%\vspace{2cm}
%{\bf VOIR SI AJOUT INDICATION LIMITATIONS SPECIFIQUES POUR CHACUNE !!! PAS FAIT LORS DU COURS ; 
%TABLE UN PEU LARGE ... }
%\vspace{2cm}

\begin{table*}
\caption{Summary of mitigating techniques. References indicate papers proposing the method or using it, but the list is not always exhaustive. For example, many analysis rely on the use of Gaussian processes and chromospheric emission indicators. (*) including convective blueshift inhibition in plages at the rotational timescale (**) including convective blueshift inhibition in plages at all timescales (***) including convective blueshift inhibition in plages at long timescales.}
\label{tab_techn}
\begin{center}
\renewcommand{\footnoterule}{}  % to avoid a line before footnotes
\begin{tabular}{l|l|l}
\hline
\multicolumn{3}{c}{Methods based on usual RV time series only}  \\ \hline
Fits of sinusoids around P$_{\rm rot}$  & spot and plage  (*) & \cite{boisse11} \\
\hspace{0.5cm}  and harmonics & &  \\
Prewhitening of the signal at Prot & spot and plage  (*) &  \cite{queloz09}  \\ 
 &  &  \cite{hatzes10}\\
Spot modeling & spots and plages (*) & \cite{moulds13} \\
 &  &  \cite{dumusque14} \\
 & &  \cite{herrero16} \\
Averaging of the  signal & granulation & \cite{dumusque11b} \\
Periodogram standardization & granulation  & \cite{sulis17b}  \\
\hline
\multicolumn{3}{c}{Methods based on activity indicators computed from the spectra}  \\ \hline
Correlation with the bisector span  & spot and plages & \cite{queloz01}  \\
 & &  \cite{desort07}\\
 & & \cite{boisse09}\\
Correlation with the chromospheric  & plages (**) & \cite{boisse09} \\ 
\hspace{0.5cm}emission & & \cite{dumusque12}\\
 & & \cite{meunier13}  \\
 & &  \cite{robertson14}  \\
 & &  \cite{rajpaul15} \\
 & & \cite{lanza16}\\
 & &  \cite{borgniet17}\\
Use of gaussian processes based &  spots and plages (*) & \cite{rajpaul15}   \\
\hspace{0.5cm}on those activity indicators & & \cite{dumusque17} \\
 & &  \cite{malavolta18} \\
 & & \cite{damasso18} \\
PCA analysis & spots and plages  & \cite{davis17} \\
Doppler imaging techniques & spots  & \cite{hebrard16} \\
\hline
\multicolumn{3}{c}{Methods based on RV time series computed from subsets of data}  \\ \hline
Produce independent time series  & plages (***) & \cite{meunier17c} \\
\hspace{0.5cm}based on different sets of lines & & \\
Use of selected lines with different  & spots and plages (**) & \cite{dumusque18}  \\
\hspace{0.5cm}sensitivity to magnetic field & &\cite{cretignier20}\\
Wavelength dependence of the signal  & spots and plages & \cite{talor18} \\ 
\hspace{0.5cm}and chromatic index& & \\
\hline
\multicolumn{3}{c}{Methods based on activity indicators obtained separately from the spectra} \\ \hline
Estimation of the RV signal using the  & spots &  \cite{aigrain12} \\
\hspace{0.5cm} photometric light curve (FF' method) & & \\
\hline
\end{tabular}
\end{center}
%\tablefoot{test}
%\tablefoot{References indicate papers proposing the method or using it, but the list is not always exhaustive. For example, many analysis rely on the use of Gaussian processes and chromospheric emission indicators. (*) including convective blueshift inhibition in plages at the rotational timescale (**) including convective blueshift inhibition in plages at all timescales (***) including convective blueshift inhibition in plages at long timescales.}
\end{table*}

%A REPORTER 
%inhibition of the convective blueshift in the same way, and to estimate its amplitude if the dynamics of the variability is large enough (cycle) and the S/N large enough plages \cite{meunier17c}
%Use of selected lines with different properties (line depth formation, wavelength, ...) which are not sensitive to the magnetic signal in the same way pspots and plages \cite{dumusque18, cretignier20}
%Wavelength dependence of the signal due to the contrast of spots and plages, with a chromatic index, if the wavelength coverage is high enough spots and plages e.g. \cite{talor18}

\begin{figure} 
\includegraphics[width=10cm]{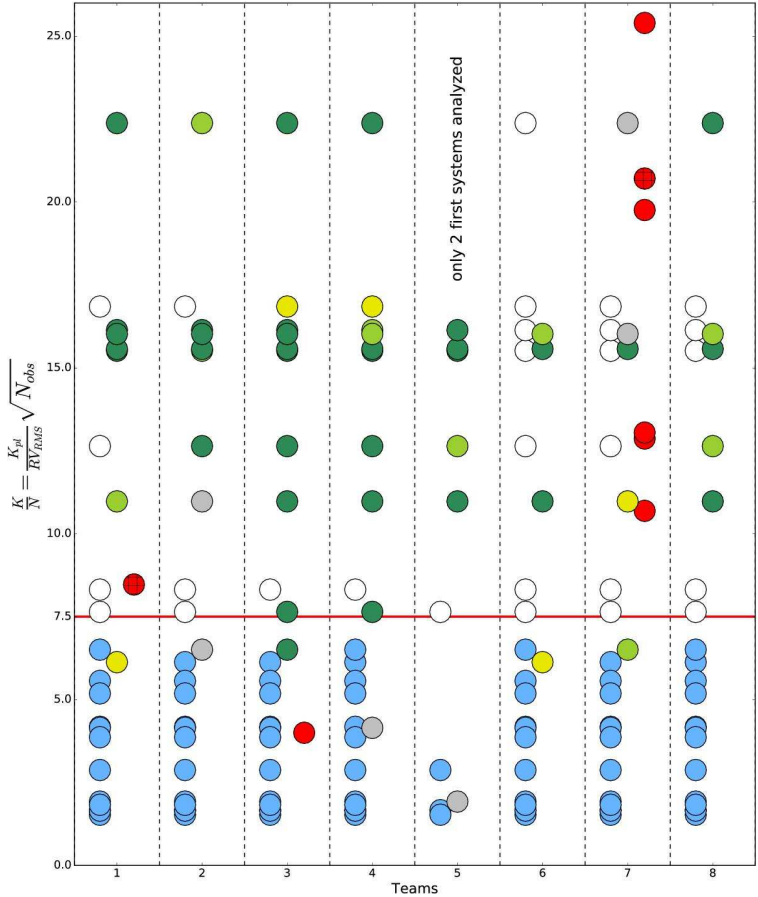} 
\caption{Results from the fitting challenge organized by \cite{dumusque17}, representing K/N (see text) for each team and all planets in five synthetic systems. The color flags indicate the status of the detection (or lack of detection): planet not detected (K/N$<$7.5 in blue, K/N$>$7.5 in white), planet properly detected (green), detected planet but wrong parameters (yellow if claimed, gray if probable), false positive or negative (red), probable detection but no planet injected (orange). From \cite{dumusque17} reproduced with permission \copyright ESO.  } 
\label{fitt_chall} 
\end{figure}

%legende d'origine : K/N ratio for all the planets present in the 5 first systems of the RV fitting challenge, in addition to the false positives and the false negatives announced for those 5 first systems. All the teams analyzed those 5 first systems, except team 5 that only looked at the first two systems, which explain why there is less dots corresponding to planetary signals. The different color flags are defined in the legend of Fig. 13 and in more details in the second paragraph of Sect. 4.2. We separate the false positives or false negatives appearing in red in two categories. Either they cannot be explained easily (plain red dots), or the activity signal at the stellar rotation period has been confused with a planetary signal despite the fact that the correct stellar rotation period was found a priori (hatched red dots). The red horizontal line corresponds to a K/N ratio of 7.5. The RV rms used to calculate K/N is the rms of the raw RVs once the best-fit of a model consisting of a linear correlation with ) plus a second order polynomial as a function of time was removed. This model allows removing the effect of magnetic cycles and any long-term drift in the RVs. We removed from this plot the 2 planets in system 11 and 12 that have an orbital period longer than 3000 days, much longer than the timespan of the data, which explain why they were not detected by any team despite their large K/N values (see Sect. 4.2.4).

All these methods significantly reduce the stellar contribution  to some level, but they never remove it completely. However, they also all have limitations and lead to crucial questions. First, since none of them removes completely the stellar signal, what is their limit in terms of performance? Second, how reliable are the residuals: do they remove part of any planetary signal, do they add any spurious signal?
There are many reasons for these limitations: 
\begin{itemize}
\item{Strong degeneracies between the numerous processes are present, which make them difficult to estimate.} 
\item{Involved processes are very stochastic, with a very complex frequency behavior of the stellar processes, and contributions at all timescales.}
\item{The models used in these methods are not perfect and do not describe with enough accuracy the processes (or do not include any physics at all), and  do not take all processes into account.}
\item{Some parameters needed in those models are not always properly estimated, for example the rotation period, which cannot be uniquely defined for a given star because of differential rotation. }
\end{itemize}

Finally, the very sparse sampling and the presence of photon noise complicate the situation.  In addition, very recent results by \cite{roy20} also show the possible contamination of RV measurement above the 10 cm/s level due to the Moon (solar light). We also note that in principle, some of these methods should provide interesting insights  on stellar activity itself (such as the rotation period, properties  about spots and plages, and on activity cycles), although they usually focus on exoplanets only and therefore on the residuals after subtraction of part of the activity contribution. 

The fitting challenge organized by \cite{dumusque16} allowed to test some of these methods (8 teams participated to the challenge) on a few complex time series, most of them synthetic  (see next section) in a blind test. The results are presented and discussed in \cite{dumusque17}. The injected planets covered a very large range in period ($\sim$0.8-3400 days). Some of these methods allowed the teams to  retrieve some planets well (mostly the techniques based on Gaussian processes) while other did poorly (no planet  or wrong planet retrieval). The same was true for the retrieval of the rotation period. They also propose a criterion defined as K/N=K$_{\rm pla}$ $\sqrt{N_{\rm obs}}$/$\sigma$, where K$_{\rm pla}$ is the amplitude of the RV planet signal (which can be converted into a mass, providing an orbital period) and $\sigma$ is the RV jitter after a correction of the RV time series using a linear correction with $\log R'_{HK}$ and second degree in time polynomial fit. As shown in Fig.~\ref{fitt_chall}, they evaluate  an approximate value of 7.5 for this criterion, above which the retrieval performance was bad and below which the performance was good in many cases. This shows that given current techniques, the retrieval of low mass planets (such as the Earth) in the habitable zone around solar type stars is not possible  \cite[as also shown by][on a large set of activity simulations using that criterion, see next section]{meunier19b}.

\subsection{Simulations}

An important  way to study in more detail the stellar processes affecting RV measurements and their impact are simulations. I define here three categories. Simulations involving a very simple magnetic configuration (typically 1-2 spots) for a given set of parameters and simulations involving complex activity (spot, plage) patterns aiming at reproducing solar-like activity time series, have been implemented to reproduce the impact of magnetic activity. The general  approach for these two categories is the following, although some of the steps can be omitted in simplified versions: at each time step, structures (spots, plages)  are defined (position, size, contrast), localized on a map or not, with spectra associated to each contribution from the surface, providing a final spectrum after integrating on the disk. This spectrum can then be analysed like an observed spectrum (after adding white noise for example, or extracting a certain sampling). A third category concerns photospheric flows. I review their objective and a few selected results below.

\subsubsection{Simulations with 1-2 spots}

The objective of simple simulations, i.e. with typically 1-2 spots, is to derive typical amplitudes and shapes for simple activity configurations, in order to understand and identify fine effects related to a single structure at the rotational timescale. When the star exhibits a simple configuration, it can also be used to fit  actual observations. Different tools have been implemented independently (list not exhaustive), such as SAFIR \cite[][]{desort07}, SOAP and SOAP2 \cite[][]{boisse12,dumusque14}, and Starsim \cite[][]{herrero16} for this purpose. 

I list here a few representative results. \cite{desort07} simulated a single dark spot (plage and convective blueshift inhibition were included later) and studied the impact of v sin i, inclination, spectral type, and spot position, as well as the relationship between RV and BIS (see Sect.~2.4). The main results are the following: the BIS does not exhibit any temporal variation if the spectral resolution is low; there are regimes with significant stellar RV variations but no BIS variations; the general behavior can be described as scaling laws depending on the instrument and spectral types; latitude and inclination strongly impacts the signal; and chromatic effects are observed as expected. They note that data inversion should be difficult to perform because there are too many parameters. \cite{boisse12} obtained similar results, but also studied the effect of latitude in more detail as well as the impact of the limb darkening function. They also performed a comparison with observations. Their model was improved in \cite{dumusque14}, with the addition of plages and in particular the inhibition of the convective blueshift. The spectra of the quiet Sun and of spots were used as inputs and they provide detailed temporal variations depending on various assumptions for those spectra. As an illustration of an application of the tool, the modeling is presented for two stars, one where a single plage is dominating ($\alpha$ Cen B, plage size of 2.4\% at a latitude of 44$^\circ$ and an inclination of 22$^\circ$) and one where a single spot is dominating (HD189733, spot size of 0.8\% at a latitude of 61$^\circ$ and an inclination of 80$^\circ$).

\subsubsection{Simulations with complex activity patterns}

The objective of complex activity patterns is also to identify and study fine effects, but for more complex configurations at the rotation period as well as on long timescales (for example taking the variability of the activity level into account). It also allows to study detectability and characterization performance for solar-like stars using a systematic analysis (such as the impact of different samplings, tests of correction methods...) and to find new diagnosis to develop new correction techniques. This has been made for the observed Sun \cite[][]{lagrange10b,meunier10a,meunier12,meunier13,makarov10b,haywood16},  for a simulated Sun \cite[][]{borgniet15} and then for other stars: for a few stars in the fitting challenge \cite[][]{dumusque16}, for spots only  \cite[][]{santos15} and for a large range of stellar parameters in a systematic way \cite[][]{meunier19,meunier19b,meunier19c}. Results from the fitting challenge are presented above (Sect.~4.1). Here I focus on the approach described in this latter series of papers.

\begin{figure} 
\includegraphics[width=13cm]{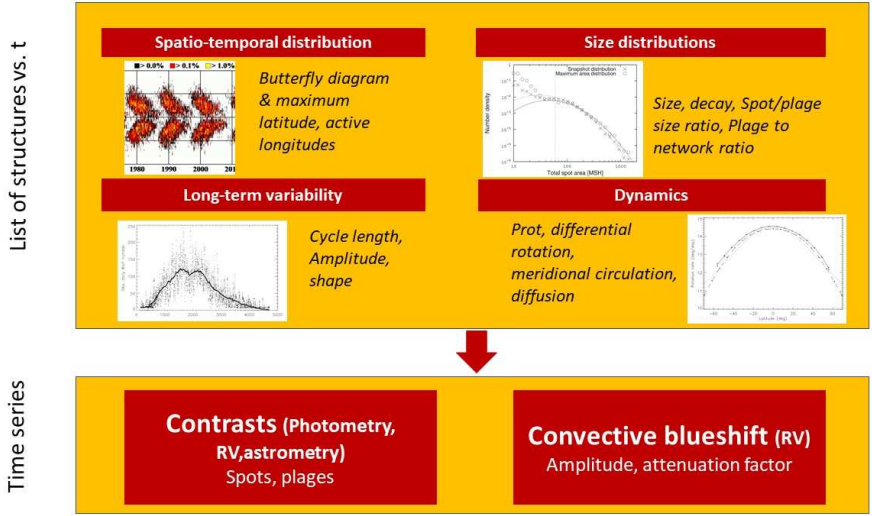} 
\caption{Principle of the simulations performed in \cite{borgniet15} for the Sun. The first step consists in generating spots and plages (position, size) at each time step based on various categories of empirical laws listed here. The second step generates the time series of observables. } 
\label{simus}
\end{figure}

The best characterised star is the Sun, and the manifestation of solar activity has been the subject of a huge effort from the solar community for decades. I first focus here on the manifestation of magnetic activity and in particular the contribution directly due to spots and plages (described in Sect.~3.2 and 3.6). The starting point of the approach was the question: if we were to observe the Sun from the outside, would we be able to detect the Earth given our current instruments and methods? The study focused on the system star-planet (or Earth-Sun) without the presence of additional more massive planets like Jupiter which would have to be taken into account for a complete solar system and which would be responsible for additional ``noise". The main principles of the approach were: 1/ use of our knowledge of the Sun and extending the analysis to other stars; 2/ simulations of complex activity patterns; 3/ systematic approach in terms of parameter space and performance evaluation; 4/ use of these simulations to find new diagnosis and improve correction methods.

A first step was performed by reconstructing the solar integrated RV due to magnetic activity using actual observed structures (spots, plages) and a model \cite[][]{lagrange10b,meunier10a}: this allowed to study the Sun seen edge-on during cycle 23 and showed that the dominant contribution came from the inhibition of the convective blueshift in plages, with a long-term amplitude of the order of 8 m/s, i.e. two orders of magnitude higher than the Earth signal. This was illustrated in Fig.~\ref{challenge}. This was confirmed with a reconstruction of the solar integrated RV from MDI/SOHO Dopplergrams \cite[][]{meunier10}, as well as from HMI/SDO Dopplergrams \cite[][]{haywood16}: this is presented in Sect.~4.4. 
A model (Fig.~\ref{simus}) was then implemented to produce realistic sets of spots and plages in the solar case, based on empirical laws such as the solar butterfly diagram, the cycle shape, size distribution, plage-to-spot ratio distribution etc. \cite[][]{borgniet15}. RV time series were then computed from the list of structures as in the reconstruction made in 2010. This model allowed to study the dependence of the variability on inclination.  At low inclinations compared to the edge-on configuration, the amplitude of the signal around the rotation period strongly decreases, although there is still some contribution present around similar scales due to the evolution and lifetime of structures. The long-term amplitude slightly decreases (due to the fact that plages, at low latitudes, cover a lower apparent proportion of the disk), but is still important (a few m/s). 
This step also allowed to be able to extrapolate this work to solar type stars, which was done in \cite{meunier19}, described below.

\begin{figure} 
\includegraphics[width=12cm]{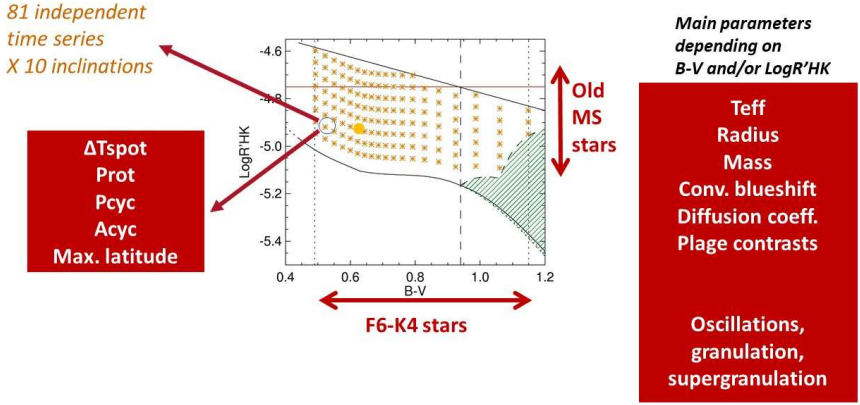} 
\caption{Range of stars covered by the simulations of \cite{meunier19}, extrapolating the solar simulation of \cite{borgniet15} described in Fig.~\ref{simus} to other stars. Parameters adapted to different stellar types and activity levels are indicated. } 
\label{extrapol}
\end{figure}

\begin{figure} 
\includegraphics[width=12.5cm]{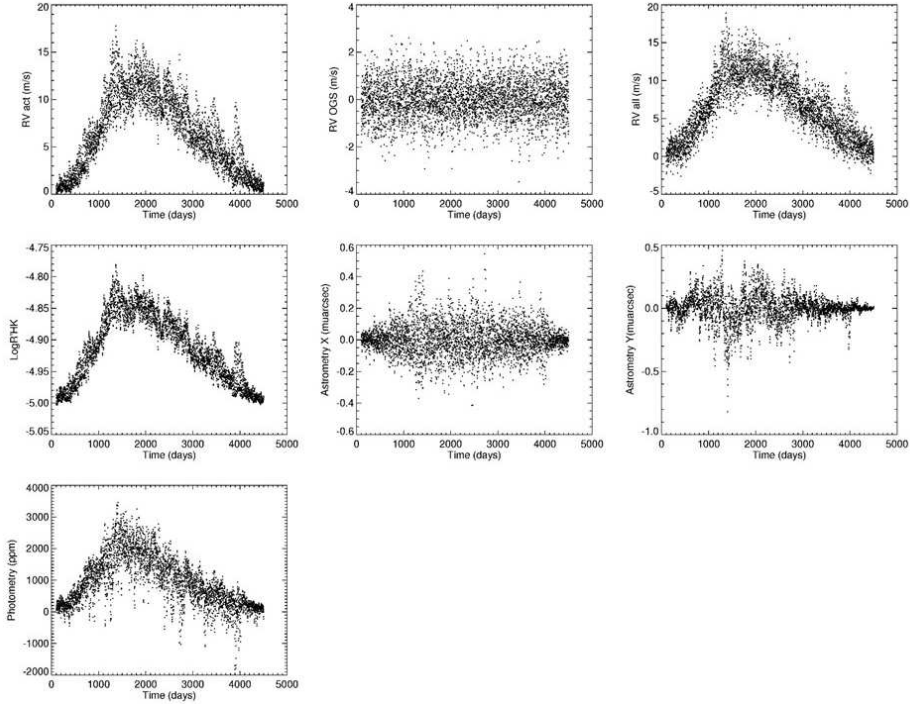} 
\caption{Example of time series produced by the simulations made in \cite{meunier19}. } 
\label{ex_serie}
\end{figure}

In \cite{meunier19}, the extrapolation to solar-like stars was limited to F6-K4 stars and old main sequence stars  \cite[plage dominated regime][]{lockwood07} to select a regime where some of the solar parameters would still be reasonable (Fig.~\ref{extrapol}).  In that domain, for each activity level and spectral type, a number of parameters are explored to cover various configurations. Rotation rates for example depend on activity level and spectral type \cite[][]{mamajek08} while the amplitude of the convective blueshift inhibition can be deduced from the analysis of a large sample of stars \cite[][]{meunier17,meunier17b}. A good knowledge of stellar activity (observational and theoretical) is necessary to perform such simulations. The parameters which are explored within a range compatible with observations for each point in Fig.~\ref{extrapol} are: rotation period, cycle period and length, spot contrast, and maximum latitude of the butterfly diagram. All laws are described and discussed in \cite{meunier19}. Other observables than RV were also produced: chromospheric emission $\log R'_{HK}$, photometry \cite[][]{meunier19d}, and astrometry \cite[][]{meunier20}. An example of time series is shown in Fig.~\ref{ex_serie} (one realisation, edge-on configuration). 

\begin{figure} 
\includegraphics[width=9cm]{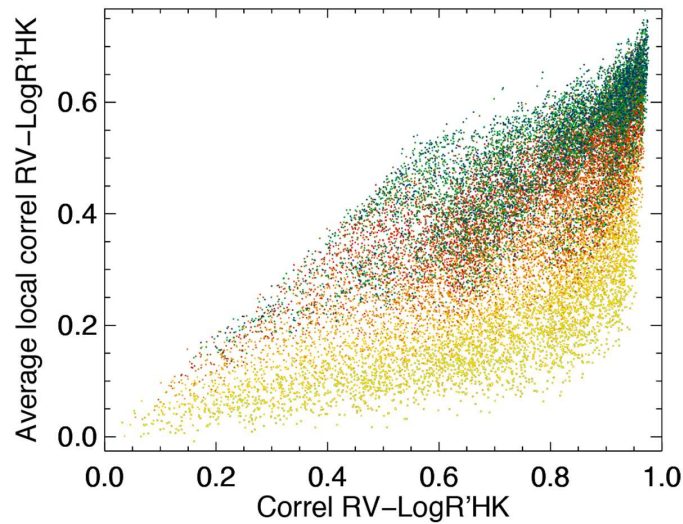} 
\caption{Average local correlation between RV and $\log R'_{HK}$ vs. the global correlation over a large sample of simulations. The color code indicates stellar inclination (from pole-on in yellow to edge-on in blue). From \cite{meunier19c} reproduced with permission \copyright ESO.}
\label{fig_correl} 
\end{figure}

The properties of the synthetic time series have been compared to other variables, for example photometry, and $\log R'_{HK}$, and to published RV jitter trends with spectral type \cite[][for example]{wright05} in \cite{meunier19b}. They show a good agreement of the trend, although the RV jitter is smaller in the simulations. In particular, they discussed the fact that given our current understanding of stellar activity, it is impossible to reproduce a jitter of a few m/s in the case of very quiet ($\log R'_{HK}$ lower than -5 for example) stars, meaning that either the uncertainties has been underestimated in published RV jitters, or that an additional contribution (not stellar, for example planets, or stellar, such as meridional circulation) are present. However, there is a good agreement of the RV versus $\log R'_{HK}$ slope with the observations of \cite{lovis11b}. In addition, Fig.~\ref{fig_correl} shows a comparison between the average local correlation (computation on small subsets of the synthetic time series, typically at the rotational timescale) between RV and $\log R'_{HK}$,  and the global one (computed over complete time series, i.e. dominated by the cycle of activity). This shows that although the long-term correlation is often close to 1, the local correlation can be much smaller. It also strongly varies for a given time series depending on the current activity level for a given time series.

A simple analysis of the RV jitters allows to estimate the order of magnitude of the detection limits for this range of stars. Using the criterion proposed by \cite{dumusque17} (see above for definition, K/N$\sim$7.5), a rough estimate of the typical mass which could be detected with current techniques can be performed. \cite{meunier19b} show that with a low number of observations (100 nights, representative of current ordinary time series), the minimum mass is of several M$_{\rm Earth}$, and only for low mass stars (the threshold is above 10 M$_{\rm Earth}$ for F stars), with maximum values of several tens of M$_{\rm Earth}$, for planets in the habitable zone. A 1 M$_{\rm Earth}$ level can only be reached for a very large number of nights  (several thousands over a long period) and low mass stars (K stars mostly). A more sophisticated approach will be the subject of a future work.

\begin{figure} 
\includegraphics[width=12.5cm]{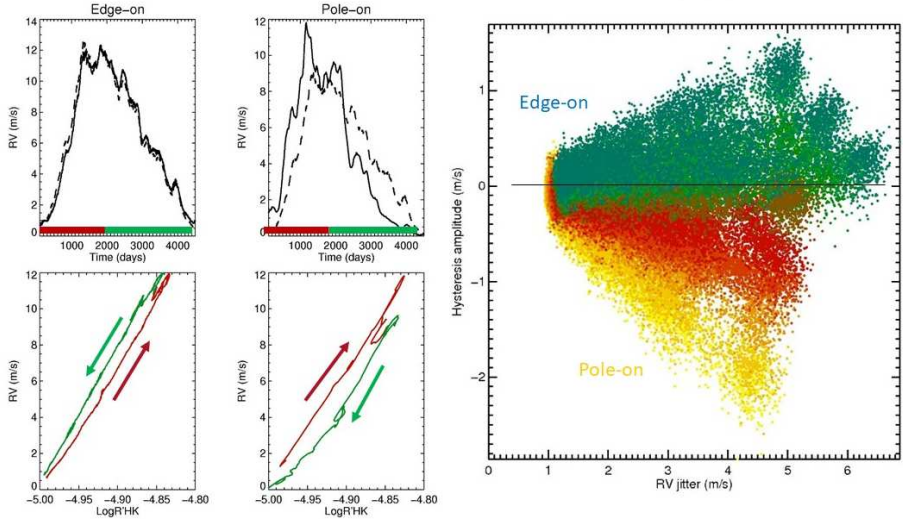} 
\caption{Left panels: example of smoothed synthetic time series (RV as a solid line and scaled $\log R'_{HK}$ as a dashed line, upper panels) and RV vs. $\log R'_{HK}$ (lower panels). The ascending phase of the cycle is in red and the descending phase in green and have a different behavior, as shown in \cite{meunier19c}. Right panel: amplitude of the hysteresis vs. RV jitter for different inclinations (from pole-on in yellow to edge-on in blue), from \cite{meunier19c} reproduced with permission \copyright ESO. }
\label{hyst_ex} 
\end{figure}

\begin{figure} 
\includegraphics[width=13cm]{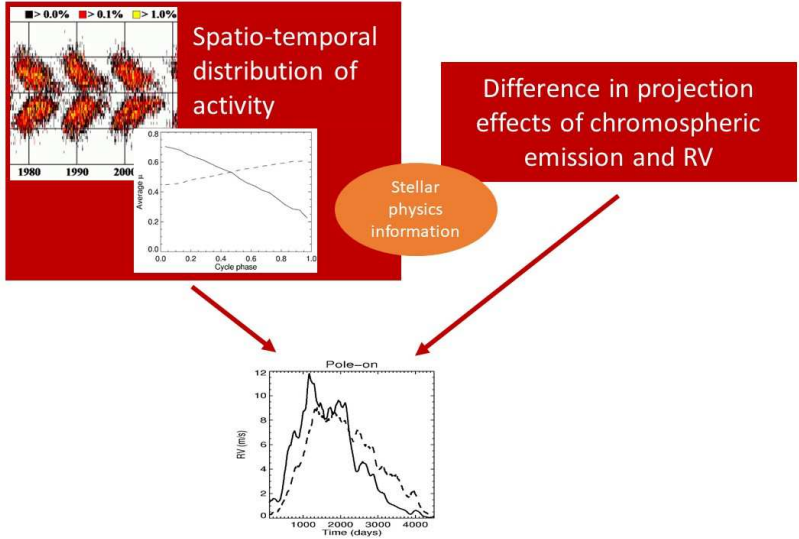} 
\caption{Physical processes leading to the hysteresis pattern observed in Fig.~\ref{hyst_ex}. The butterfly diagram (NASA/NSSTC/HATHAWAY) leads to a change in average $\mu$ with time, which depends on inclination. The average $\mu$ vs. cycle phase and RV vs. time are from \cite{meunier19c} reproduced with permission \copyright ESO.. }
\label{hyst_principe} 
\end{figure}

\begin{figure} 
\includegraphics[width=8cm]{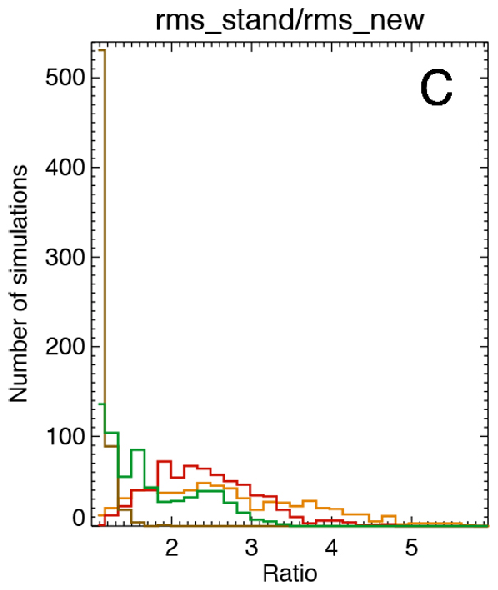} 
\caption{Gain in rms (on the smooth synthetic time series) for four different inclinations, G2 stars, with a standard correction method (linear correlation in $\log R'_{HK}$) and the new method taking the combined effects of the butterfly diagram and projection effects into account. From \cite{meunier19c} reproduced with permission \copyright ESO. }
\label{hyst_res} 
\end{figure}

Finally, \cite{meunier19c} have shown that this type of approach can be used to improve correction techniques. A widely used method to remove part of the stellar signal is to apply a linear correlation between RV and $\log R'_{HK}$ \cite[e.g.][]{boisse09,dumusque12,robertson14,rajpaul15,lanza16,diaz16,borgniet17}. It is well known that it is not perfect \cite[e.g.][]{meunier12}, but it allows to remove a significant part of the signal when the inhibition of the convective blueshift is dominant, because both are strongly correlated with the plage filling factor. These simulations put a departure from the linear correlation into evidence, and are useful to understand why there is such a departure and improve the correction techniques. This is illustrated in Fig.~\ref{hyst_ex}: the long-term variation shows a different relationship during the ascending phase of the cycle and during the descending phase. This is due to two effects related to the geometry, the dynamo processes and surface processes (Fig.~\ref{hyst_principe}): 1/ the butterfly diagram (activity pattern as a function of time and latitude, shown in the upper left panel of Fig.~\ref{hyst_principe}) leads to a different average $\mu$ over time: a decrease when the star is seen pole-on and an increase (to a lesser extent) when the star is seen edge-on, because the activity pattern moves from mid-latitudes to the equator during the cycle; 2/ RV and chromospheric emission suffer from different projection effects. This leads to a distortion of the RV time series with respect to the $\log R'_{HK}$ time series, which could be corrected by using a correlation with the $\log R'_{HK}$ times a function describing the contribution of these two effects. The application of this technique led to a significant improvement of the long-term residuals in those simulations (Fig.~\ref{hyst_res}).

\subsubsection{Simulations of the small or large scale dynamics}

The simulation of photospheric flows has taken different forms in the literature. For meridional circulation, it has been performed using direct observations of the flows, i.e. the observed latitudinal profiles over time \cite[][]{makarov10b,meunier20c}. For granulation and supergranulation, some simulations have also been based on properties deduced from hydrodynamical simulations for granulation and/or supergranulation by \cite{meunier15} using the simulations of \cite{rieutord02}, directly from those hydrodynamical simulations \cite{cegla16,cegla18,cegla19b,sulis20}, and from the Harvey's law \cite{meunier19e,meunier20b}. 

For stellar meridional circulation \cite{meunier20c} extrapolated the solar results using the results of \cite{brun17} from MHD simulations. Some results have been presented in Sect.~3.4, 3.5 and 3.8 and are not repeated here. This type of approach allows to model synthetic RV time series, which can be used to characterize the amplitude of the contributions due to these processes, but also, as for magnetic activity, to test performance (detection rates, false positives) and to improve correction techniques. 

\subsection{Stellar observations: simultaneous campaigns}

Another way to understand better RV time series is to perform simultaneous RV observations with other observables. Spectroscopic indicators as described above (such as line shape indicators and $\log R'_{HK}$ measurements) are widely used, but it is important to note that they are either noisy (bisector shape for example) or representative of one of the contributions only: chromospheric emission is mostly sensitive to plages and in the end representative of the convective blueshift inhibition contribution. The use of simultaneous photometry, which is sensitive to the contribution to the contrast of spots and plages is therefore particularly interesting to constrain better the different processes at play, altough it does not model all of them. \cite{aigrain12} has for example proposed the FF' method, to reproduce the RV signal due to spots (mostly for 1-2 spot configurations) from the photometry. Simultaneous measurements have therefore been done \cite[e.g.][]{cloutier17,lopez19} for a few stars. Such coordinated campaigns, aiming at covering well the rotation period for a good sampling, are very difficult to implement, because ground-based photometric observations are much noisier than from space and are subject to meteorological conditions, but they will be useful in the future to provide a complementary approach. The model derived from photometry is also incomplete due to the presence of other process in RV, as described in Sect.~3 \cite[e.g.][]{haywood16}.

\subsection{Observations of the Sun as a star}

Finally, as already pointed out in previous sections, our knowledge of the Sun is crucial to progress on these issues. Traditionally, there have been a lack of long-term stable integrated RV for the Sun, since most studies focused on observations with a good spatial resolution. \cite{mcmillan93} obtained an upper limit of 4 m/s using deep lines (which are not very sensitive to the convective blueshift contribution at small wavelengths), while \cite{DP94} observed a peak-to-peak amplitude of 30 m/s in the infrared (2.3 $\mu$m) and \cite{jimenez86}  obtained a large variability (30 m/s on the long-term, 20 m/s on the short-term) using the K 7699 \AA$\:$ line. Later on, the reconstruction of the solar RV using Dopplergrams from MDI/SOHO by \cite{meunier10}, whose objective was to reproduce the contribution of active regions (and mostly the convective blueshift inhibition due to the noise level), was compatible with simulations as described above, with a long-term amplitude of the order of 8 m/s, which confirmed the reconstructions made from observed structures by \cite{meunier10a} and later by \cite{haywood16} from HMI/SDO. Other solar observations were then performed using an indirect approach by \cite{lanza16} (observations of Jupiter satellites, asteroids, and the Moon), with a variability compatible with those previous results. 

A significant step has been made with the advent of ``Sun as a star" observations using stellar high-performance stabilized spectrograph, the solar light being supplied to the spectrograph through a coelostat and an integrated sphere. Even though this reproduces stellar observations, some adaptation in the data processing had also to be done due to the finite solar apparent diameter, which led to some spurious effects (differential effects due to the atmosphere for example). Such a device has been implemented on HARPS-N for the last 4 years \cite[][]{dumusque15,phillips16,collier19}: the Sun is observed about 6 hours per day with a 5 minute cadence. It has also been implemented more recently on HARPS in La Silla, or on Expres (Lowell observatory) and several similar projects are on-going. Some results have already been obtained by \cite{milbourne19}. The comparison with solar reconstruction from Dopplergrams and other indicators such as the unsigned magnetic field is also very promising \cite[][]{haywood20}.

\section{Conclusion}

Stellar activity contributes to RV in a complex manner because of the large number of processes, covering a large range of time scales but also similar orders of magnitude. There is also a large amount of degeneracies between the different contributions. A large diversity is observed among stars of different spectral types and ages. It is however complementary to other observables (in particular because some contributions can be seen only in RV) and it is also crucial to understand stellar activity observed in RV given the challenge it constitutes to observe low mass planets. Techniques to push the limits to be able to detect low mass planets (like the Earth), and to characterise their mass, including at long orbital periods (habitable zone), will therefore have to be improved and be based on our physical knowledge of stellar activity to be able to control the residuals after correction. Recent studies also showed that there was still to learn about the Sun to fully understand its integrated RV. 
The development of highly performant instruments such as Espresso is critical to improve our understanding, because they allow to develop more sophisticated techniques, for example dependent on spectral lines for example (requesting high S/N even after computation on a subset of lines). A large wavelength coverage in the optical and infrared is also extremely  promising. 

One important conclusion however if we want to be able to push the limits toward very low mass planets in the habitable zone is the fact that a huge number of points ($>$ 1000) will be necessary to better take stellar activity into account, which could constrain future facilities. The fitting challenge organized by X. Dumusque \cite[][]{dumusque16,dumusque17} or the complex activity pattern simulations \cite[][and following papers]{meunier19} also show that given the challenge and the complexity of stellar activity (especially its high level of stochasticity and the lack of independent activity indicators for certain contributions), it will be necessary to develop new techniques, and probably to combine them to be able to obtain robust results in terms of detection. 

In this context, it is also important to recall  the other factors contributing to the difficulties of studying stellar activity in RV, in particular the usually very sparse and irregular sampling (apart for the Sun) and the fact that it could be superimposed to other contributions, and in particular undetected planets (or massive planets removed but with some uncertainties on the parameters): stellar activity strongly affects exoplanet detectability and characterisation, but the presence of exoplanets can also in turn impact stellar RVs!

Finally, this topic is crucial for the preparation of the PLATO mission \cite[launched in 2026][]{rauer14} and the exoplanet science that will be performed during the mission, since the primary targets are Earth mass planets in the habitable zone around solar type stars (using photometric transits): the RV follow-up of such detections will be complicated by the fact that current techniques do not yet allow to reach very precise mass estimation in such conditions, at least for Earth mass planets. A complete review of processes and tools is given in Watson et al. (in prep).

%The InfraRed Astronomy Satellite (IRAS; Beichman {\em et al.\/} \cite{Bei}) ushered in the era of space-based infrared astronomy in a dramatic fashion, revealing a stunningly rich infrared sky, unanticipated from the bits of infrared data previously and heroically collected from the ground...IRAS had a profound influence on astronomy in general, not just the infrared, because it represented such avery large gain in sensitivity and spatial coverage, comparable perhaps togoing from attempting visual astronomy in daylight to observing in a darknight (Beichman \cite{ref1987}; Soifer \etal \cite{so1987}).  Anothersignificant factor in this influence was the  disseminationof data products from IRAS, includingsource catalogs, image atlases and skybrightness estimates, generated with great care and well characterized anddocumented as to reliability, completeness and other statisticalattributes.

%%-----------------------------
%%      your bibliography
%%-----------------------------
%\begin{thebibliography}{99}
%\bibitem[1994]{alref1} Aalto, S. \etal\  1994, A\&A, 286, 365.
%%%% Using \cite{Bei} in the text
%\bibitem[1986]{Bei} Beichman, C.A., Neugebauer, G., Habing,
 %  H., Clegg, P.E. \& Chester, T.C. 1988, editors, {\it ``IRAS Catalogs and
 %  Atlases: Explanatory Supplement''}, NASA RP-1190 (Washington: NASA)
%\bibitem[1987]{ref1987} Beichman, C.A. 1987, ARA\&A, 25, 521
%\bibitem[1987]{so1987} Soifer, B.T., Houck, J.R. and Neugebauer, G. 1987, ARAA, 25, 187
%\end{thebibliography}

\bibliographystyle{astron}
\bibliography{proceeding_meunier}

\end{document}